\documentstyle[psfig]{mn2e}

\newcommand{\HI}{H\,{\sc i}}
\newcommand{\vhel}{$v_{\rm hel}$}
\newcommand{\wfi}{$w_{\rm 50}$}
\newcommand{\wtw}{$w_{\rm 20}$}
\newcommand{\vCMB}{$v_{\rm CMB}$}
\newcommand{\FHI}{$F_{\rm HI}$}
\newcommand{\MHI}{$M_{\rm HI}$}
\newcommand{\Mdyn}{$M_{\rm dyn}$}
\newcommand{\msun}{$M_{\odot}$}
\newcommand{\lsun}{$L_{\odot}$}
\newcommand{\kms}{~km\,s$^{-1}$}     
\newcommand{\gal}{HIZOA\,J0836--43}
\newcommand{\GG}{2MASX J08363600--4337556}
\newcommand{\GGG}{2MASX J08370723--4339137}

  \title[A Massive Spiral Galaxy in the Zone of Avoidance]
      {A Massive Spiral Galaxy in the Zone of Avoidance\thanks{The 
       observations were obtained with the Australia Telescope which 
       is funded by the Commonwealth of Australia for operations as 
       a National Facility managed by CSIRO.}}

\author[J. L. ~Donley et al.]
       {J. L. Donley$^{1,2}$, 
	B. S. Koribalski$^1$,
        L. Staveley-Smith$^1$, 
        R. C. Kraan-Korteweg$^{3,4}$, \and
        A. Schr\"oder$^5$,
        P. A. Henning$^6$ \\
        $^1$Australia Telescope National Facility, CSIRO, 
            P.O. Box 76, Epping, NSW 1710, Australia\\
        $^2$Steward Observatory, University of Arizona, 933 N. Cherry Ave.,
	    Tucson, AZ 85721, USA \\
        $^3$Department of Astronomy, University of Cape Town, 
	    Private Bag X3, Rondebosch, 7701, South Africa \\
        $^4$Depto. de Astronom\'{i}a, Universidad de Guanajuato, 
            Apdo. Postal 144, Guanajuato, GTO 36000, Mexico \\
        $^5$Department of Physics and Astronomy, University of Leicester,
	       Leicester LE1 7RH, UK \\
        $^6$Institute for Astrophysics, University of New Mexico, 
	       800 Yale Blvd., NE, Albuquerque, NM 87131, USA\\
}

\date{Accepted 2006 April 4. Received 2006 March 31; in original form 2005 Dec 7}

\begin{document}

\maketitle

\begin{abstract}

We report the discovery of a very \HI-massive disk galaxy, \gal, at a
velocity of $v_{\rm hel}$ = 10689\kms, corresponding to a distance of
148 Mpc (assuming $H_0=75$\kms\,Mpc$^{-1}$). It was found during the
course of a systematic \HI\ survey of the southern Zone of Avoidance
($|b| \le 5\degr$) with the multibeam system at the 64\,m Parkes radio
telescope. Follow-up observations with the Australia Telescope Compact
Array (ATCA) reveal an extended \HI\ disk. We derive an \HI\ mass of
$7.5 \times 10^{10}$ \msun. Using the \HI\ radius, we estimate a total
dynamical mass of $1.4 \times 10^{12}$ \msun, similar to the most
massive known disk galaxies such as Malin\,1. \gal\ lies deep in the
Zone of Avoidance ($\ell, b = 262\fdg48, -1\fdg64$) where the optical
extinction is very high, $A_{\rm B} = 9\fm8$. However, in the
near-infrared wavebands, where the extinction is considerably lower,
\gal\ is clearly detected by both DENIS and 2MASS. Deep AAT
near-infrared ($K_{\rm s}$ and $H$-band) images show that \gal\ is an
inclined disk galaxy with a prominent bulge (scale length 2.5\arcsec\
or 1.7 kpc), and an extended disk (scale length 7\arcsec\ or 4.7 kpc)
which can be traced along the major axis out to a radius of 20\arcsec\
or 13.4 kpc (at 20 mag\,arcsec$^{-2}$ in $K_{\rm s}$). The \HI\ disk
is much more extended, having a radius of 66 kpc at 1
\msun\,pc$^{-2}$. Detections in the radio continuum at 1.4 GHz and at
60 \micron\ (IRAS) are consistent with \gal\ forming stars at a rate
of $\sim$35 \msun\,yr$^{-1}$. We compare the properties of \gal\ with
those of the most \HI-massive galaxies currently known, UGC\,4288,
UGC\,1752 and Malin\,1, all of which are classified as giant low
surface brightness galaxies.
\end{abstract}


\section{Introduction} 

\begin{figure*} 
\begin{tabular}{c}
   \mbox{\psfig{file=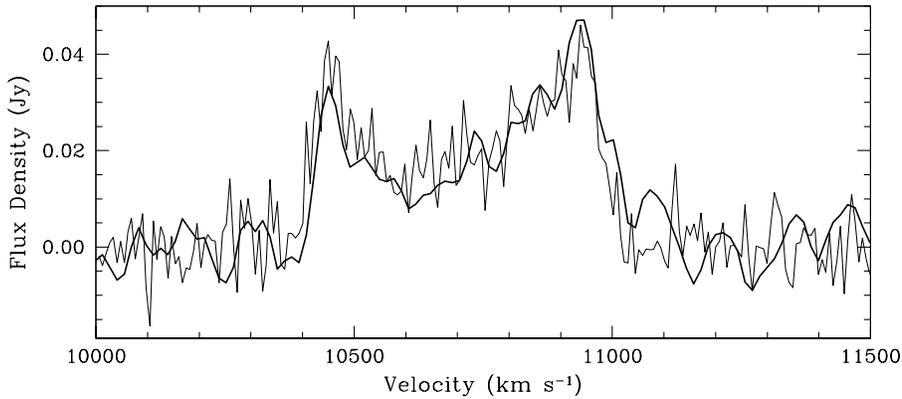,width=12cm}}
\end{tabular}
\caption{Global \HI\ profile of the galaxy \gal\ as obtained with the 
   64-m Parkes telescope (thick line) and the ATCA 750D-array (thin
   line).}
\end{figure*}

Hierarchical models of structure formation in the early Universe
predict that massive galaxies begin to form from the merging of
low-mass systems at early times and that the mass and luminosity
functions for galaxies evolve in a self-similar manner (Press \&
Schechter 1974; Cole et al. 2000). Such models (Kauffmann, White \&
Guiderdoni 1993; Klypin et al. 1997) predict large numbers of old
low-mass systems and relatively small numbers of young high-mass
systems. However, the detailed match between observations and
predictions of the mass function is poor. Two well-documented
discrepancies are (1) the small number of low-mass systems observed in
the local Universe (Davies, Sabatini \& Roberts 2004) and (2) the
large number of massive, distant red galaxies, implying a significant
density of massive galaxies at early times (Foerster Schreiber et
al. 2004).

These discrepancies most probably arise because of the fundamental
role of gas physics (supernova and AGN feedback, etc.) in determining
the star-formation efficiency in halos of different mass (Kay et
al. 2002). Indeed, a recent attempt to incorporate gas physics by the
application of a suite of semi-analytic models to the Millennium run
(Croton et al. 2006, Springel et al. 2005) has been remarkably
successful in reproducing the exponentially decreasing volume density
of objects more luminous than a characteristic value $L^*$, as well as
the observation that massive galaxies tend to contain older stars than
low-mass systems.

Due to the sharp decline in the number of galaxies with $L>L^*$ (or
$M>M^*$), large samples are needed to constrain the high-luminosity
and high-mass ends of the galaxy luminosity and mass functions.  In
the optical, such a sample is provided by the 2dF Galaxy Redshift
Survey, from which Norberg et al. (2002) find $M^*_{b_j} = -19\fm0$
($H_0=75$\kms\,Mpc$^{-1}$) and a slight tendency for an overabundance
of massive galaxies relative to a Schechter fit (although they comment
that magnitude errors are mostly able to account for this
deviation). Similar results are obtained from the Sloan Digital Sky
Survey (Blanton et al. 2003).

To sample the high-mass end of the \HI\ mass function, blind \HI\
surveys such as the \HI\ Parkes All-Sky Survey (HIPASS) are required.
HIPASS is a systematic \HI\ survey of the southern sky within the
radial velocity range of $-1200 < cz < 12700$\kms\ (Meyer et
al. 2004).  Zwaan et al. (2005) find a characteristic mass in neutral
hydrogen gas of $M^*_{\rm HI} = 6.3
\times 10^9$ \msun\ ($H_0=75$\kms\,Mpc$^{-1}$), using a catalogue of
4315 \HI\ galaxies from HIPASS. The mass function remains
ill-constrained at the high-mass end, however, due to the increasing
sparsity of high \HI-massive galaxies.

We report here on the discovery and detailed properties of the most
massive galaxy in the \HI\ Zone of Avoidance Survey (HIZOA), a deep
Parkes \HI\ multibeam survey in the obscured region along the Galactic
Plane (\S 2) (see Donley et al. 2005, Kraan-Korteweg et al. 2005, and
Henning, Kraan-Korteweg, \& Staveley-Smith 2005 for preliminary
results). This galaxy, \gal, is also the most \HI-massive galaxy in
the southern and northern-extension HIPASS catalogues, which consist
of 4315 and 1003 \HI\ sources, respectively (Meyer et al. 2004, Wong
et al. 2005). At the mass of \gal, $M_{\rm HI} = 7.5 \times 10^{10}$
\msun, the volume density is $1.4 \times
10^{-9}$~Mpc$^{-3}$~dex$^{-1}$, a factor of $10^6$ times smaller than
the corresponding value at $M^*_{\rm HI}$.

A study of massive gaseous (and stellar) systems is of considerable
interest in assessing whether the details of various galaxy formation
scenarios, particularly the processes of merging and star-formation,
remain viable under the most extreme conditions. The \HI-massive
galaxy to have received most attention is Malin\,1, whose \HI\ mass is
reported to be somewhere in the range $5 - 10\times 10^{10}$ \msun\
(Bothun et al. 1987; Pickering et al. 1997; Matthews et
al. 2001)\footnote{Calibration uncertainties at the redshift of
Malin\,1 seem to be responsible for the large uncertainties in
published mass estimates.}. Malin\,1's high \HI\ mass but low optical
surface brightness is of interest in that it raises the possibility of
a cosmologically significant amount of gas being locked up in such
systems. However, such a population is not known to exist at the
sensitivities of existing \HI\ surveys (Rosenberg \& Schneider 2002;
Zwaan et al. 2003).

The radio and infrared properties of the newly-discovered galaxy are
detailed in \S 3. In \S 4, we look at the \HI\ and dynamical mass of
\gal, its star-formation rate, morphology, location on the
Tully-Fisher relation, and environment. A comparison with other
\HI-massive galaxies is given in \S 5, followed by our conclusions in
\S 6.

\section{Discovery of HIZOA J0836--43 with the Parkes telescope} 

\begin{figure*}
\begin{tabular}{cc}
   \mbox{\psfig{file=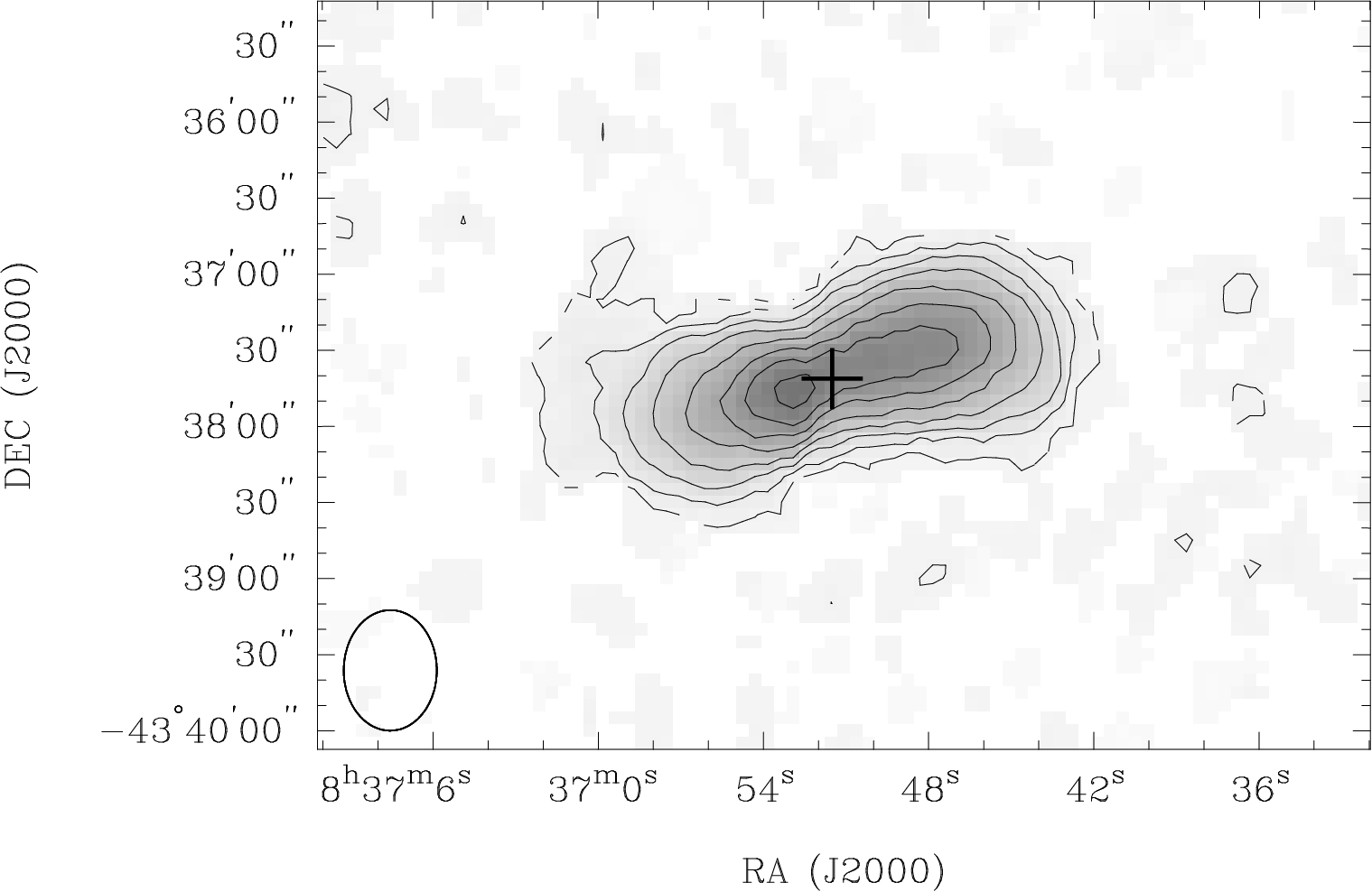,width=8cm,angle=-90}} &
   \mbox{\psfig{file=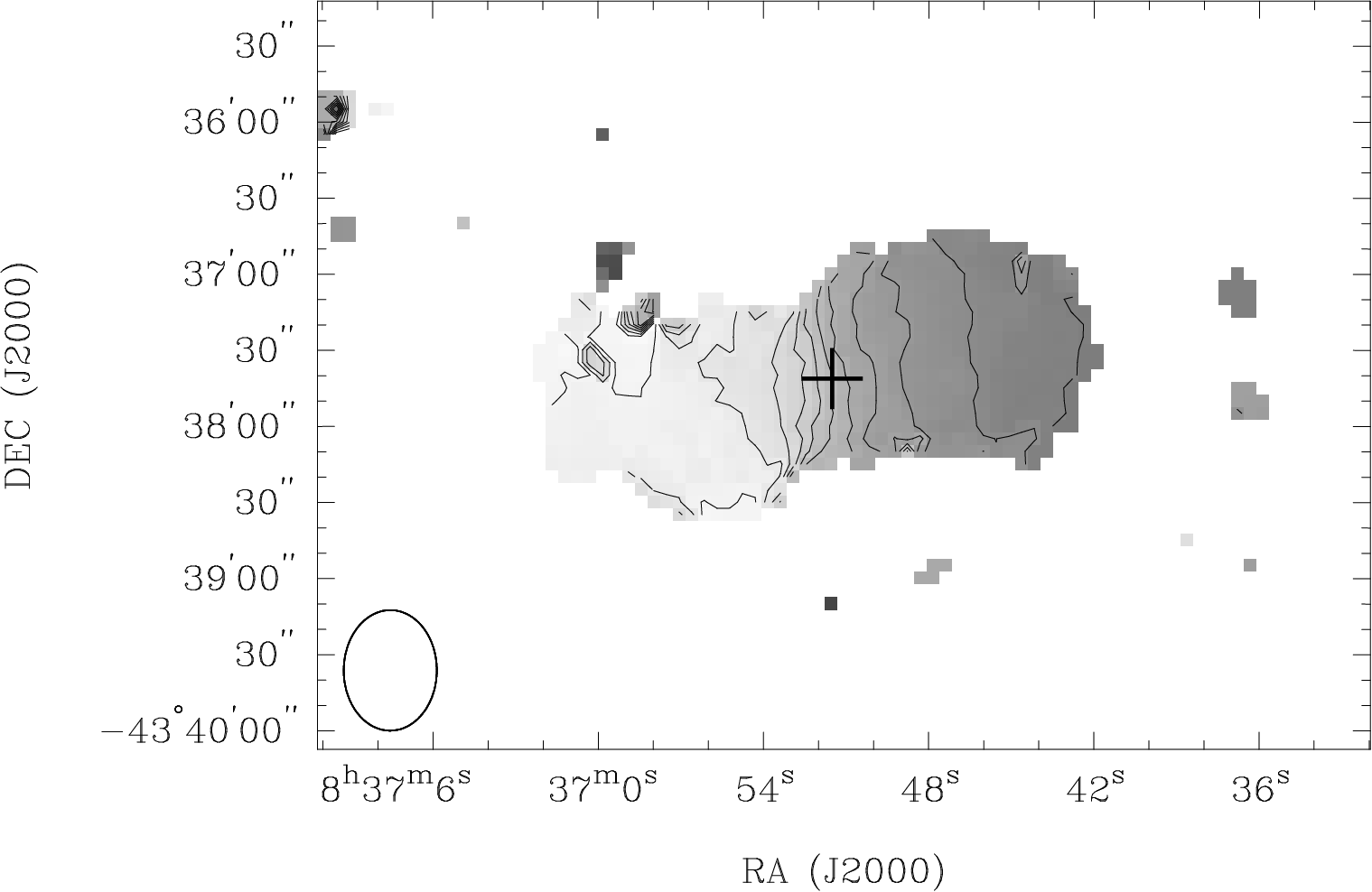,width=8cm,angle=-90}} \\
   \mbox{\psfig{file=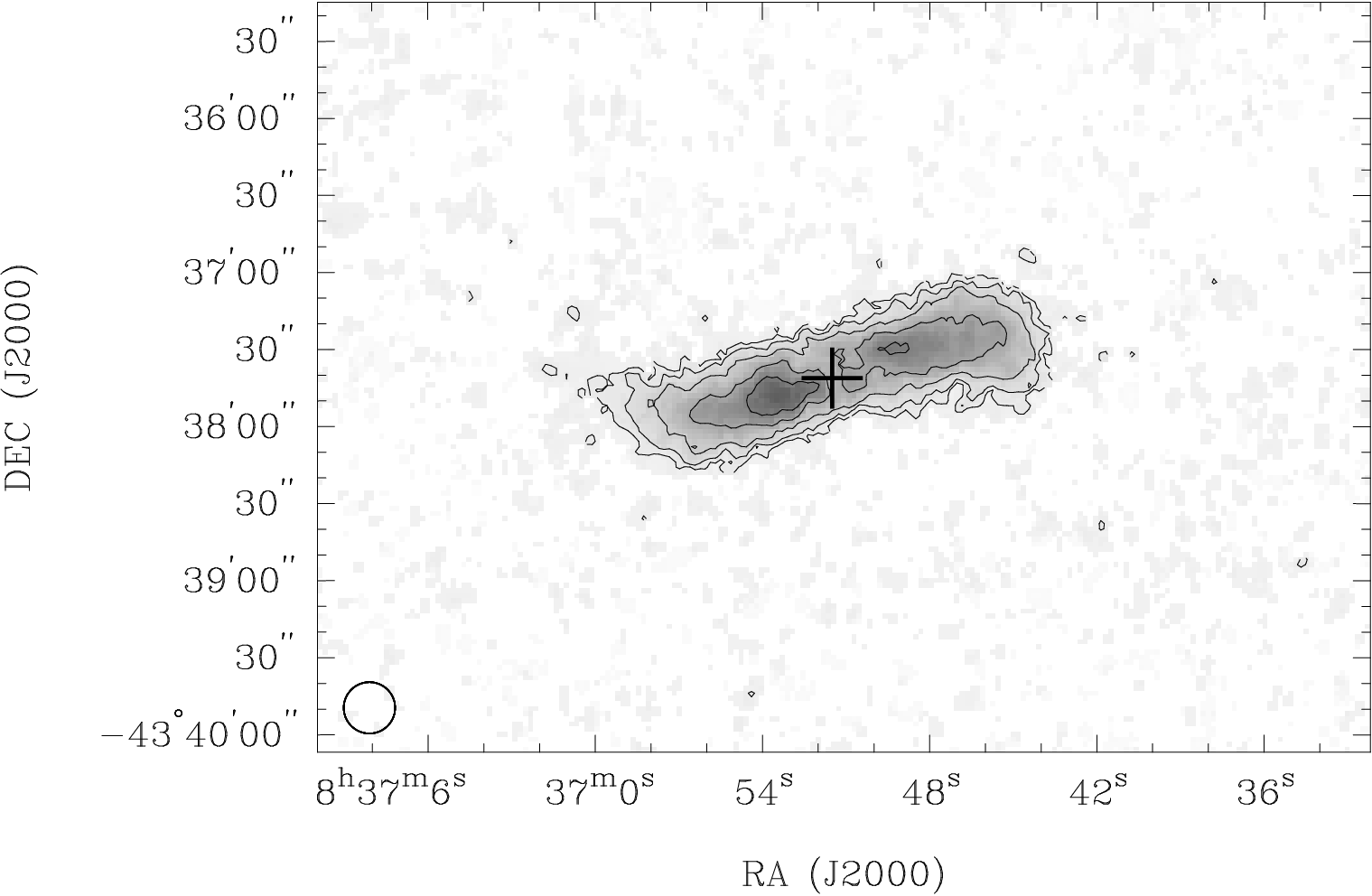,width=8cm,angle=-90}} &
   \mbox{\psfig{file=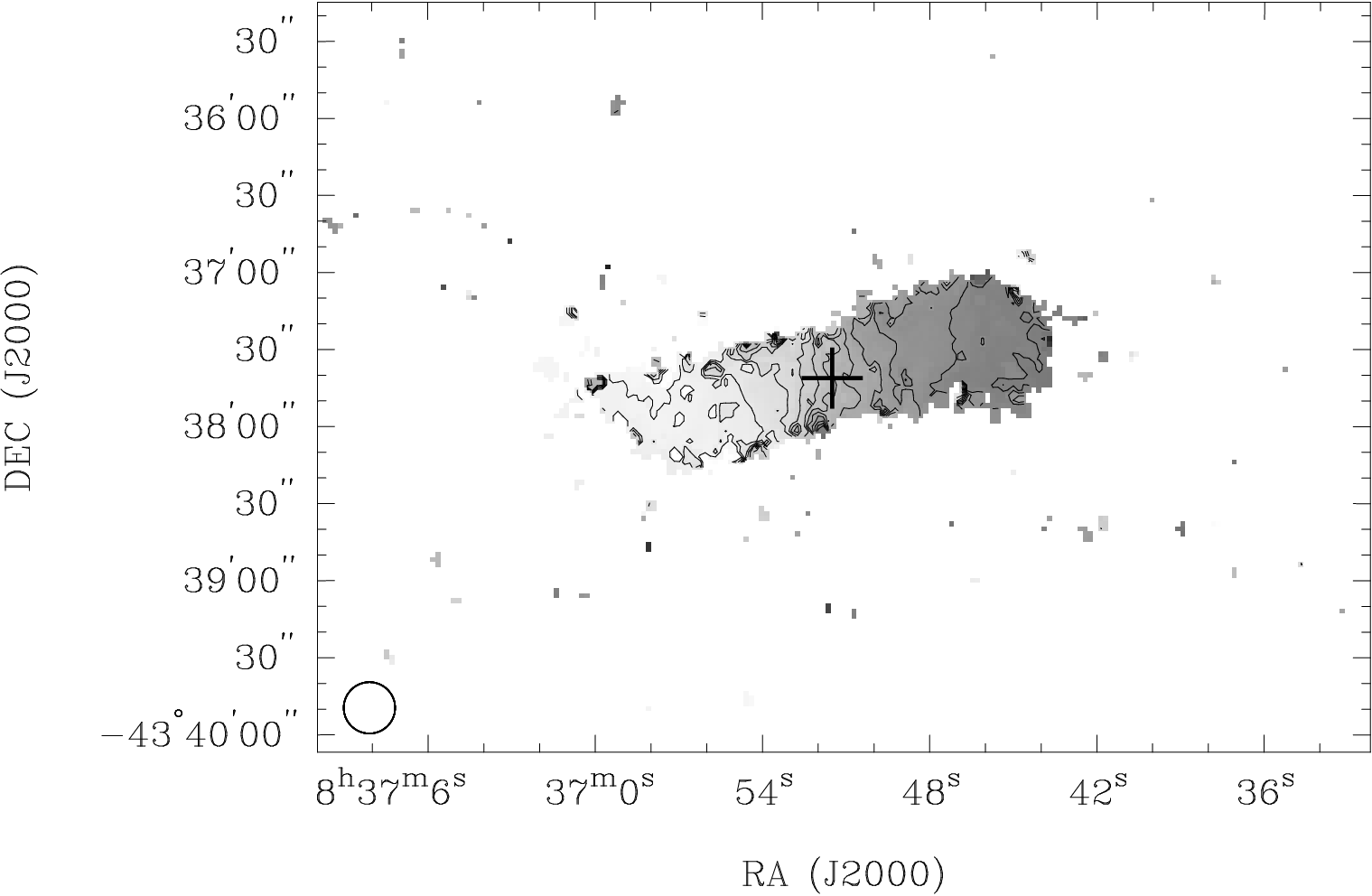,width=8cm,angle=-90}} \\
\end{tabular}
\caption{\HI\ distribution (left) and mean \HI\ velocity field (right) of 
   the galaxy \gal. Contour levels are 0.1, 0.3, 0.6, 1, 1.5, 2.0,
   2.5, and 3.0 Jy\,beam$^{-1}$\kms\ (top left; 1 Jy\,beam$^{-1}$\kms\
   equals an \HI\ column density of $6.8 \times 10^{20}$~cm$^{-2}$),
   0.1, 0.2, 0.4, 0.8, 1.2, and 1.6 Jy\,beam$^{-1}$\kms\ (bottom left;
   1 Jy\,beam$^{-1}$\kms\ equals an \HI\ column density of $3.0 \times
   10^{21}$~cm$^{-2}$), and 10450 to 10950\kms\ in steps of 50\kms\
   (right). The synthesized beam is shown in the bottom left corner
   of each frame ($47.5\arcsec \times 36.7\arcsec$ on the top, and
   20\arcsec\ on the bottom). The cross marks the galaxy centre as
   determined from the 20-cm radio continuum (see Table~2).}
\end{figure*}

The galaxy \gal\ was first detected in the deep \HI\ Zone of Avoidance
(ZOA) survey made with the Parkes 64-m telescope between March 1997
and June 2002 (see Kraan-Korteweg et al. 2005).  This blind survey
utilised the Parkes 21-cm multibeam receiver, an array of 13 beams
each with two orthogonal linear polarisations (for a detailed
description see Staveley-Smith et al. 1996). The data were calibrated
and gridded using the {\em aips++} programs {\sc livedata} and {\sc
gridzilla}, respectively; after gridding the beam width is
15\farcm5. The data reduction is nearly identical to that of HIPASS
(see Barnes et al. 2001, Henning et al.  in preparation). The \HI\
Zone of Avoidance survey covers the Galactic longitude range $52\degr
< \ell < 196\degr$, comprising the southern Milky Way (Henning et
al. in preparation) plus a northern extension (Donley et al. 2005),
for the most opaque part of the ZOA, i.e. the latitude range $-5\degr
< b < 5\degr$.  With a correlator bandwidth of 64 MHz divided into
1024 channels and an instantaneous velocity coverage of $-1200 < cz
<12700$\kms, the velocity spacing per channel is 13.2\kms, or 26\kms\
after Hanning smoothing. The approximate total integration time is
2100 sec\,beam$^{-1}$, resulting in an rms of 6 mJy\,beam$^{-1}$ in
large parts of the survey. Results of a shallow subset with an rms of
15 mJy\,beam$^{-1}$, the ZOA \HI-shallow survey (HIZSS), are given in
Henning et al. (2000).

While visually inspecting the data cube centred on $\ell = 264\degr$,
an exceptionally strong signal ($S_{\rm peak}$ = 47 mJy\,beam$^{-1}$;
see Fig.~1) for its very high velocity ($v_{\rm hel} \approx
10700$\kms) was initially identified at $\alpha,\delta$(J2000) =
$08^{\rm h}\,36.9^{\rm m}$, --43\degr\,38\arcmin\ ($\pm$3\arcmin) or
$\ell = 262\fdg5, b = -1\fdg6$.  The detected \HI\ source, named \gal,
remains unresolved in the Parkes data.

As seen in Fig.~1, the \HI\ signal of \gal\ is exceptionally broad
with a velocity width of $\Delta v \approx 600$\kms. Using the Parkes
data we measure an \HI\ flux density of 13.2 Jy\kms\ corresponding to
an \HI\ mass of \MHI\ = $7 \times 10^{10}$ \msun\ (assuming $D$ = 148
Mpc, see Section~3.1.1).  This same signal was later independently
identified in the recently released \HI\ catalogue (HICAT, Meyer et
al. 2004) obtained from HIPASS. It is listed there as HIPASS
J0836--43. The \HI\ parameters extracted from the respective surveys
are given in Table~1.

Detailed investigation of \gal\ is extremely difficult due to its
location behind the Milky Way. According to the DIRBE/IRAS dust
extinction maps of Schlegel et al. (1998) the $B$-band extinction at
the position of \gal\ is $A_{\rm B} = 9\fm8$, although we caution that
the data in the Galactic plane are not yet well calibrated.  Not
unexpectedly, this galaxy has no known optical counterpart, and
inspection of the DSS images reveal only a very faint R-band
counterpart.  The corresponding extinction in the near-infrared (NIR)
$I$, $J$, $H$, and $K$ bands is much lower: $A_{\rm I} = 4\fm4$,
$A_{\rm J} = 2\fm2$, $A_{\rm H} = 1\fm3$, and $A_{\rm K} = 0\fm8$,
respectively. DENIS ($IJK$; Epchtein 1997) and 2MASS ($JHK$; Jarrett
et al. 2000) reveal two and three possible counterparts, respectively,
in all the wavebands, 2MASX J08365157--4337407, 2MASX
J08370723--4339137 and 2MASX J08363600--4337556. As discussed in
\S3.1.1 and \S3.2, the counterpart to \gal\ is 2MASX
J08365157--4337407, a galaxy with an inclination of $\sim 66$\degr.

\section{Radio and infrared results} 

\begin{figure*} 
\begin{tabular}{c}
   \mbox{\psfig{file=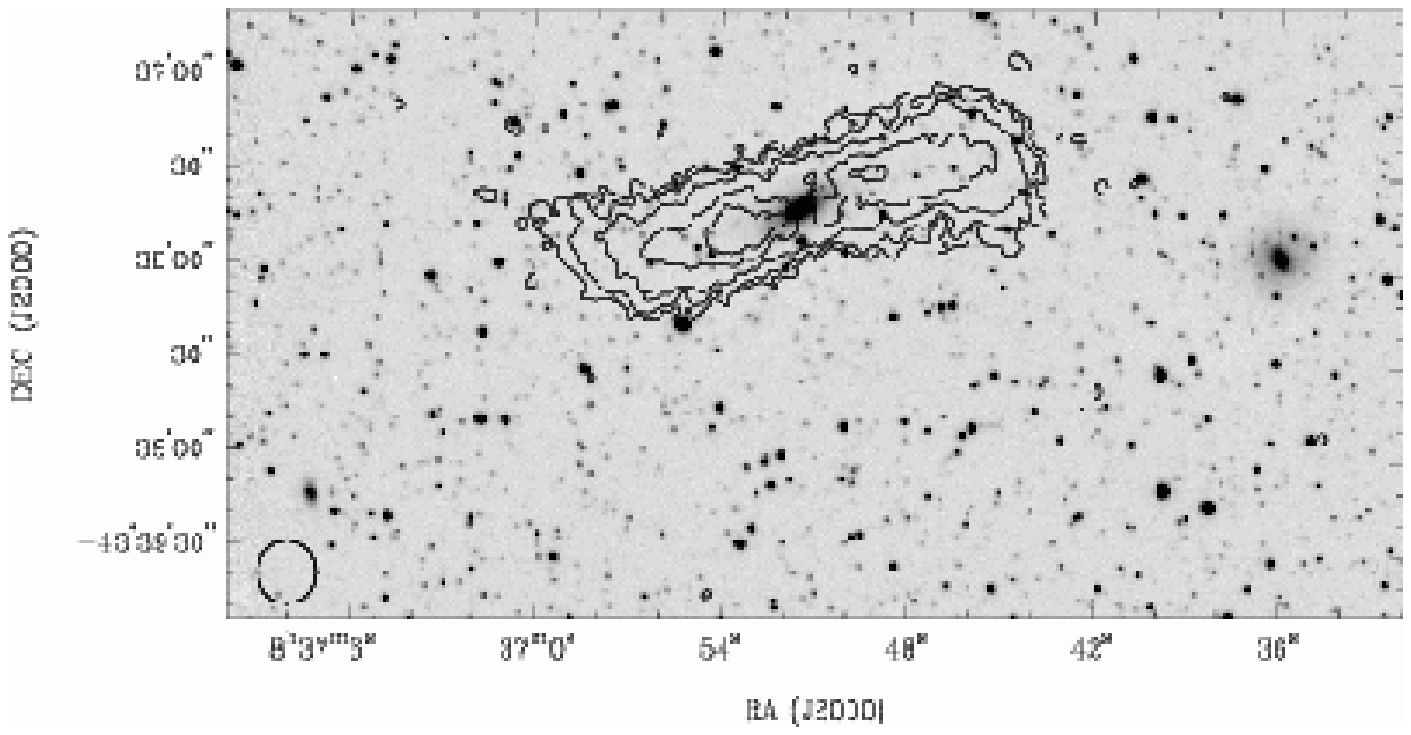,width=15.2cm,angle=0}} \\
   \mbox{\psfig{file=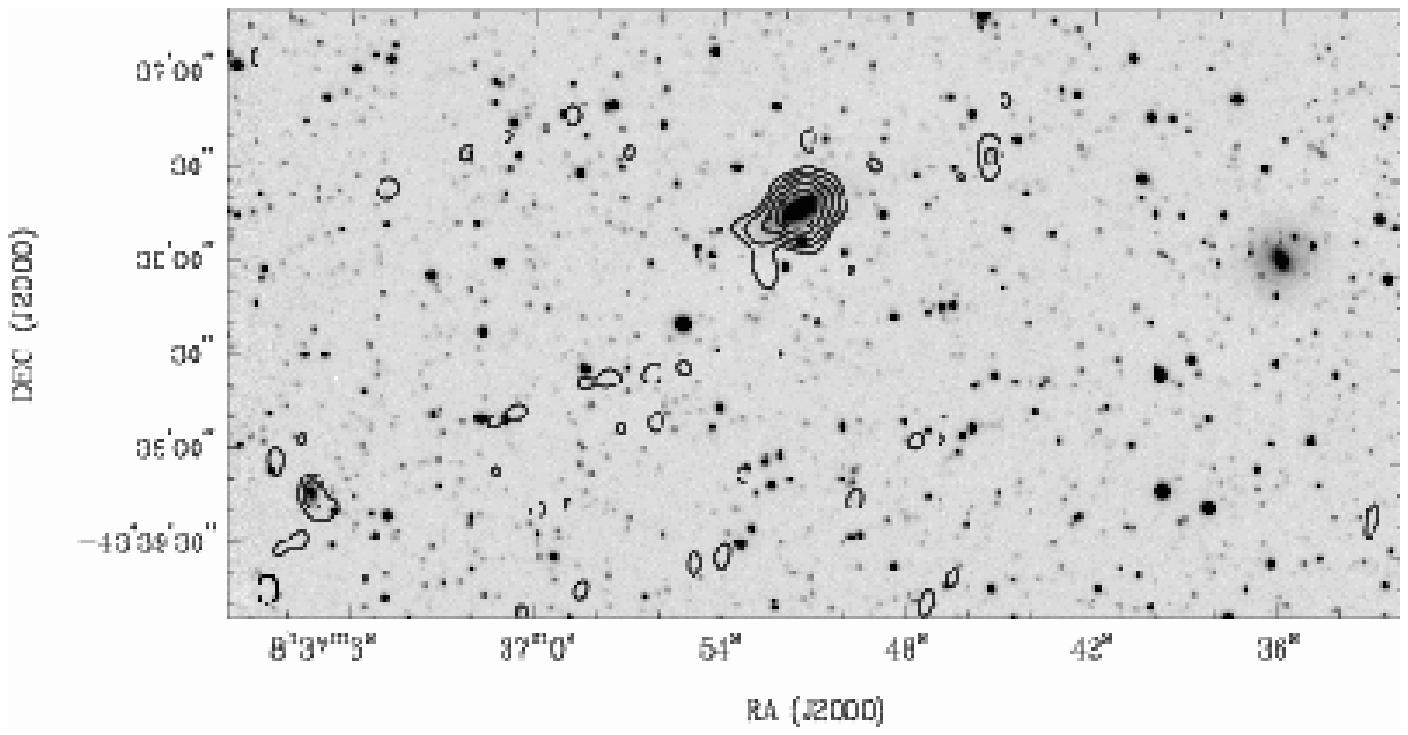,width=15.2cm,angle=0}} \\
\end{tabular}
\caption{{\bf Top:} \HI\ distribution of \gal\ (contours) at a resolution of
   20\arcsec\ overlaid onto the AAT $K_{\rm s}$-band image
   (greyscale). The contour levels are the same as in Fig.~2 (bottom
   left).  {\bf Bottom:} 20-cm radio continuum image of \gal\
   (contours) overlaid onto the same $K$-band image. The contour
   levels are --0.26, 0.26, 0.52, 1.04, 2.08, and 4.16
   mJy\,beam$^{-1}$; the synthesized beam is $8.7\arcsec \times
   6.5\arcsec$. The displayed region was chosen to also include the
   two neighbouring galaxies, one of which is detected in the radio
   continuum with an \HI\ flux density of $\sim 1$~mJy.}
\end{figure*}

\subsection{ATCA radio observations} 

\HI\ line and 20-cm radio continuum observations of the galaxy \gal\
were obtained with the Australia Telescope Compact Array (ATCA) in the
750D and the 1.5D configurations. The observations took place on 2003
February 10 and November 28, respectively, with a total integration
time of $\sim 2 \times 12$ hours. For the \HI\ line observations, we
used a centre frequency of 1371 MHz and a bandwidth of 8 MHz, divided
into 512 channels. This resulted in a channel width of 3.4\kms\ and a
velocity resolution of 4.1\kms. The wide-band 20-cm radio continuum
observations were centred on 1384 MHz with a bandwidth of 128 MHz,
divided into 32 channels. We used PKS B1934--638 as primary calibrator
and PKS B0823--500 as secondary calibrator. Their flux densities are
15.0 Jy and 5.5 Jy at 1378 MHz, respectively. Data reduction was
carried out with the {\sc miriad} software package using standard
procedures.

Various \HI\ line data cubes were made to investigate the neutral
hydrogen distribution and kinematics of the galaxy \gal. The best
results were obtained by using `natural' weighting of the combined
{\em uv}-data sets. Including all baselines resulted in an angular
resolution of $18.9\arcsec \times 18.8\arcsec$; we used a restoring
beam of 20\arcsec\ and a channel width of 10\kms. By excluding the
longest baselines (i.e. all baselines to antenna 6) we obtain an
angular resolution of $47.5\arcsec \times 36.7\arcsec$.  We measure an
rms per channel of 1.2 mJy~beam$^{-1}$ and 1.4 mJy~beam$^{-1}$,
respectively. The resulting \HI\ distribution and mean \HI\ velocity
field of \gal\ for both angular resolutions are shown in Fig.~2.

Radio continuum images were made using `robust' (r = 0) and `uniform'
weighting of the combined {\em uv}-data, resulting in angular
resolutions of $10.3\arcsec \times 9.0\arcsec$ and $8.7\arcsec \times
6.5\arcsec$, respectively. The measured rms is $\sim$0.1
mJy\,beam$^{-1}$. We find a continuum source at the position of \gal\
that is slightly extended along a position angle of $PA$ = 125\degr\
and that has an integrated 20-cm flux density of $\sim$24 mJy.

\begin{figure} 
   \mbox{\psfig{file=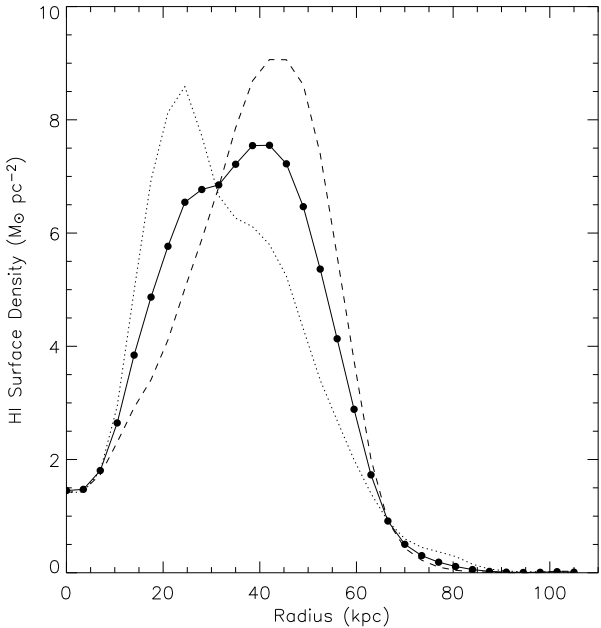,width=8cm}}
\caption{Radial \HI\ density profile of \gal, obtained with the program
   {\sc radial} in the software package {\sc gipsy}. The solid line
   (and filled circles) gives the average profile, whereas dotted and
   dashed lines give the \HI\ profile of the approaching and receding
   sides, respectively.}
\end{figure}

\subsubsection{Global HI properties} 

The global \HI\ spectra of \gal\ (see Fig.~1) were analysed using the
{\sc miriad} program {\sc mbspect}. Because the galaxy was unresolved
in the Parkes \HI\ data, its spectral profile was formed by weighting
the flux according to the beam parameters. In the ATCA \HI\ data,
\gal\ is clearly resolved and consequently the spectral profile was
taken to be the sum of the flux in a conservative box of $3\farcm8
\times 2\farcm2$, centred on the galaxy. For both the Parkes and 
ATCA \HI\ data, first-order baselines were fit to the spectra and the
total \HI\ flux density was then determined by integrating over the
galaxy profile.

The systemic velocity was taken to be the midpoint of the spectral
profile at the 50\% \HI\ peak flux level. From a heliocentric velocity
of \vhel\ = 10689\kms\ ($z$ = 0.036), we calculate a velocity in the
Cosmic Microwave Background (CMB) frame of $v_{\rm CMB}$ = 10928\kms,
where the CMB vector was taken from Fixsen et al. (1996).  Adopting a
Hubble constant of $H_0$ = 75\kms\,Mpc$^{-1}$ and a cosmology with
$\Omega_{\rm m}$ = 0.3 and $\Omega_{\rm \lambda}$ = 0.7, we derive a
cosmological luminosity distance of $D$ = 148 Mpc (see Table~2). At
this distance, 1\arcmin\ corresponds to 40.1 kpc.

The 20\% and 50\% velocity widths were measured using a
width-maximizing technique.  The measured \HI\ properties for both
observations are given in Table~1, where they are compared with
equivalent measurements from HICAT (Meyer et al. 2004).  For the
remainder of the paper, we adopt the ATCA velocity widths of $w_{20} =
610\pm9$\kms\ and $w_{50} = 566\pm6$\kms. Uncertainties in the
integrated flux density and velocity measurements were derived using
the formulas given in Koribalski et al. (2004).

Fig.~3 shows the $K_{\rm s}$-band image obtained with the AAT (see \S
3.2 for further details) overlaid onto the \HI\ distribution (upper
panel) and continuum distribution (lower panel) as obtained from the
high-resolution ATCA \HI\ data cube. Both the \HI\ and the 20~cm
continuum measurements identify the galaxy 2MASX~J08365157--4337407 as
the counterpart to \gal.  The other two galaxies,
2MASX~J08363600--4337556 and 2MASX~J08370723--4339137, lie at
distances of 2\farcm8 and 3\farcm2 from \gal, respectively.

\subsubsection{HI radial density profile} 

\begin{figure*}
\mbox{\psfig{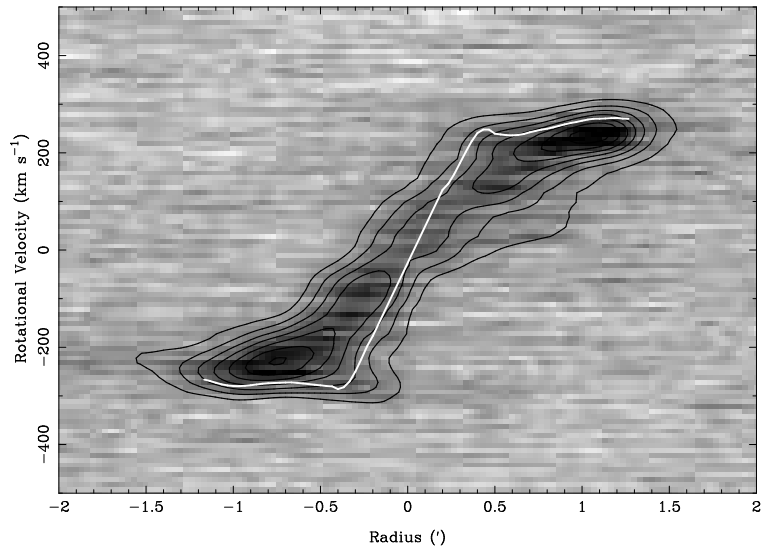}}
\caption{\HI\ position-velocity ($pv$) diagram along the major axis 
  ($PA$ = 285\degr) of \gal\ using the combined \HI\ data set with an
  angular resolution of 20\arcsec. The white line gives the \HI\
  rotation curve as shown in Fig.~8. Contours represent intensity
  levels of 2--8$\sigma$, where $\sigma = 1.14$~mJy~beam$^{-1}$, and
  are measured after smoothing the $pv$-diagram by a Gaussian kernel
  of width 2.}
\end{figure*}

\begin{figure*}
\begin{tabular}{cc}
   \mbox{\psfig{file=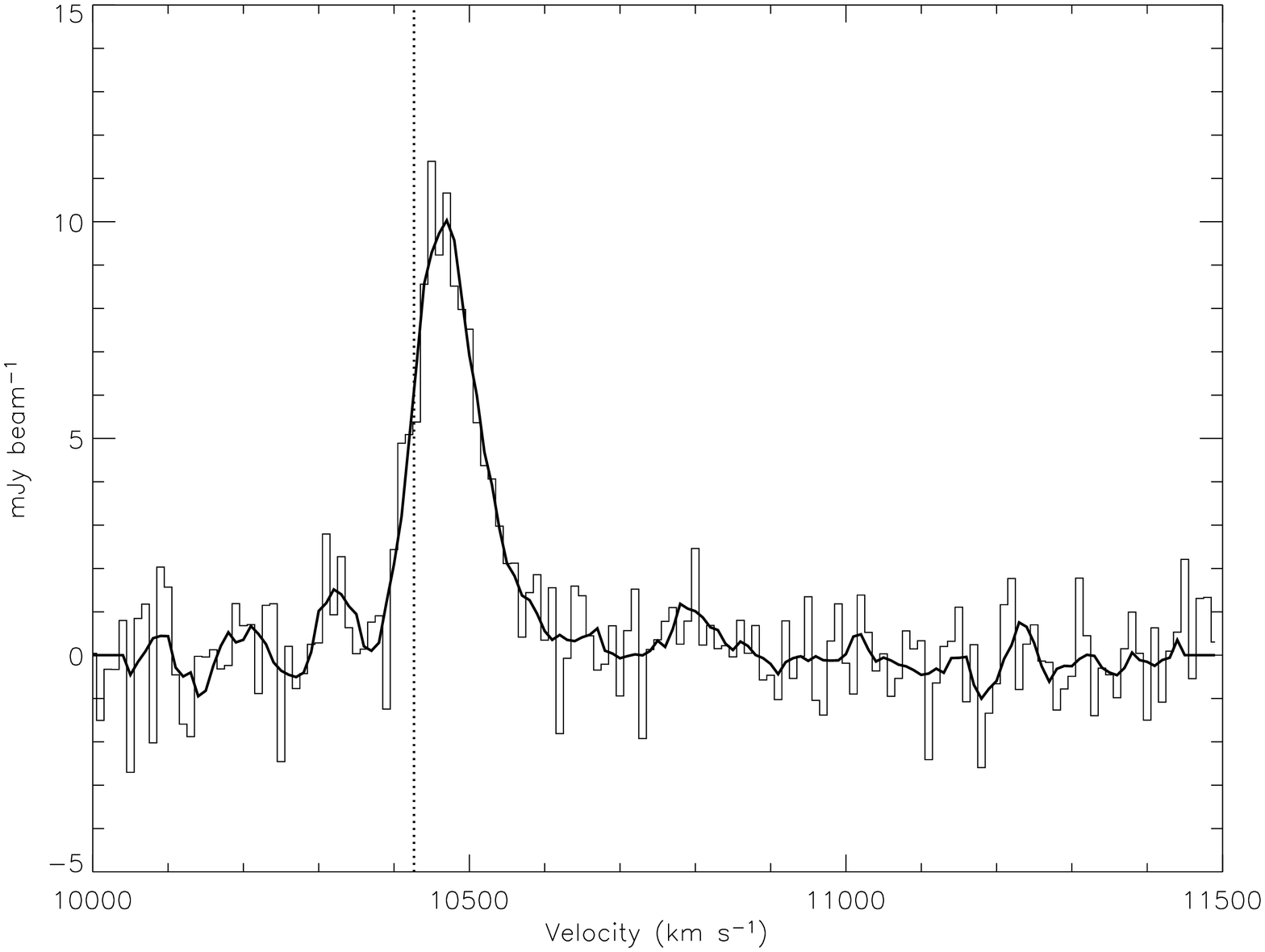,width=8cm,angle=90}} &
   \mbox{\psfig{file=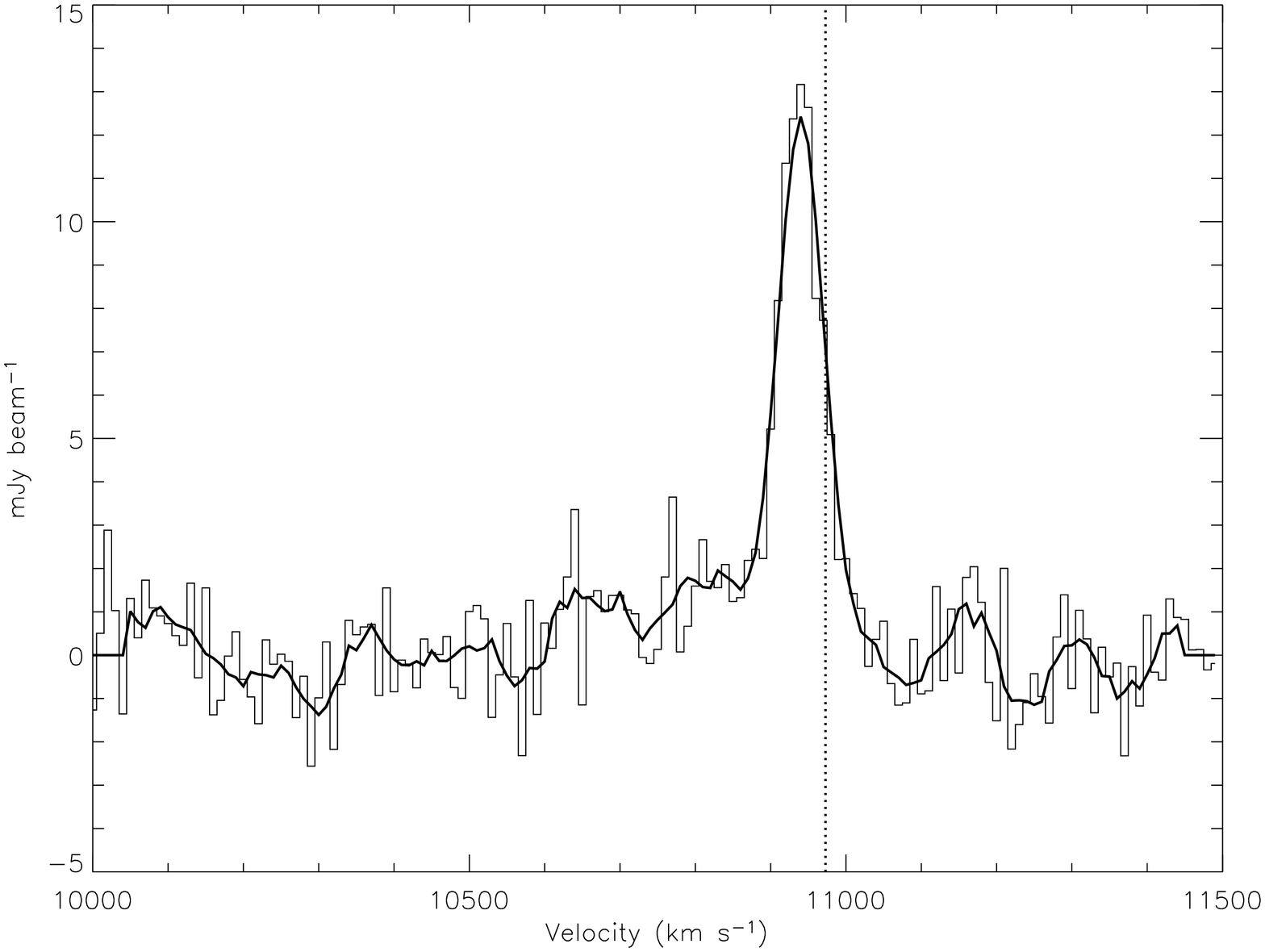,width=8cm,angle=90}} \\
\end{tabular}
\caption{Sample \HI\ velocity profiles of \gal\ obtained from the major-axis 
  $pv$-diagram at offsets of --0.9\arcmin\ and 1.0\arcmin. Dotted lines 
  represent the rotation velocity measured for each slice. Both the 
  unsmoothed and the Savitzky-Golay smoothed profiles are shown for 
  comparison.}
\end{figure*}

We measure the \HI\ radial density profile from the \HI\ total
intensity map of \gal, using the iterative method of Lucy (1974) as
applied by Warmels (1988) and implemented in the program {\sc radial}
in the software package {\sc gipsy} (van der Hulst et
al. 1992). Because this method assumes that the \HI\ has a planar and
axisymmetric distribution, it is limited in its detection of
structures such as spiral arms and warps. However, for highly-inclined
galaxies observed with low angular resolution, such as \gal, this
method provides the most accurate means of measuring the surface
density distribution. Another advantage of the Lucy method is that the
surface density is measured directly; we need not assume a given model
to which the data are fit. In addition, the density profile can be
measured independently for the two sides of the galaxy.

The \HI\ radial density profile for \gal\ as derived from the low
resolution \HI\ distribution (Fig.~2, left) is given in Fig.~4. The
density profiles of both the approaching and receding sides of the
galaxy are shown, as is the average distribution. The centre of the
galaxy was taken to be the position of the radio continuum source (see
Table~2).  The radial density profile shows a central depression, and
peaks at a surface density of 8.6~\msun\,pc$^{-2}$ and
9.1~\msun\,pc$^{-2}$ at radii of 25 kpc and 45 kpc, on the approaching
and receding sides of the galaxy, respectively.  The \HI\ radius,
measured at the position where the average radial density of the \HI\
profile drops to 1~\msun\,pc$^{-2}$, is $\sim$66 kpc.

\subsubsection{HI rotation curve} 

\begin{figure}
\mbox{\psfig{file=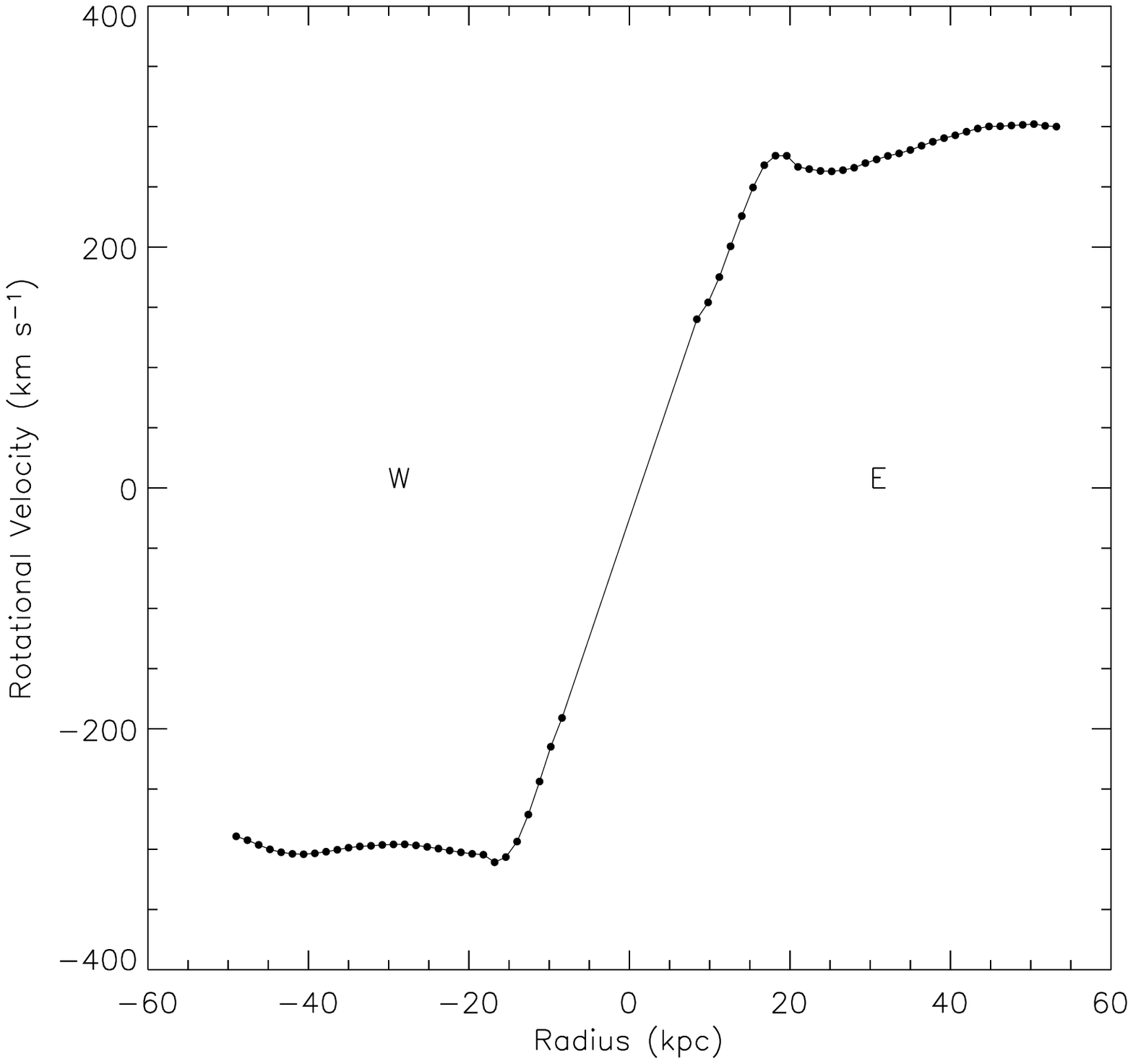,width=8cm}}
\caption{\HI\ rotation curve of HIZOA J0836--43 derived using the 
   envelope-tracing method.}
\end{figure}

The \HI\ rotation curve was measured using the envelope-tracing method
(Sofue 1996, 1997), a method known to provide reliable rotation curves
for highly inclined galaxies (Olling 1996). The rotation velocity is
given by:
\begin{equation}
V_{\rm rot} = (V_{\rm t} - V_{\rm sys})/{\rm sin}i - (\sigma_{\rm
obs}^{2} + \sigma_{\rm ISM}^{2})^{1/2}
\end{equation}
where $\sigma_{\rm obs}$ is the velocity resolution and $\sigma_{\rm
ISM}$ is the velocity dispersion of the interstellar gas. We adopt a
velocity dispersion of 7\kms\ (Stark \& Brand 1989; Malhotra
1994). The terminal velocity, $V_{\rm t}$, is the velocity at which
the intensity becomes:
\begin{equation}
I_{\rm t} = [(0.2 I_{\rm max})^{2} + I_{3\sigma}^{2}]^{1/2}
\end{equation}
where $I_{\rm max}$ and $I_{3\sigma}$ are the maximum intensity and
$3\sigma$ intensity levels, respectively.

To measure $V_{\rm t}$, we first used the {\sc karma} package {\sc
kpvslice} to create a position-velocity ($pv$) diagram along the major
axis of the galaxy (using $PA$ = 285\degr), shown in Fig.~5. We then
plotted as a function of velocity the intensity at a given radius
along the major axis. Because of the relatively low S/N, we smoothed
the profiles using a Savitzky-Golay smoothing filter with a width of
10 pixels (Press et al. 1992) before measuring $I_{\rm max}$ and the
velocity at which $I$ = $I_{\rm t}$. Fig.~6 illustrates the effect of
this smoothing on two of the velocity profiles, as well as the
resulting rotation velocity measured for each slice.  After measuring
the rotation velocity at each offset, we smoothed the rotation curve
by the beam size. Because the image resolution limits the reliability
of the central rotation curve derived using the envelope-tracing
method (Sofue 1996), we excluded from our rotation curve measurement a
central region equal to the beam diameter. We then connected the two
sides of the curve with a line, although we caution that this shape is
not necessarily representative of the true central rotation curve.

The inclination-corrected rotation curve is shown in Fig.~7; it is
also overlaid onto the $pv$-diagram shown in Fig.~5.  The rotation
velocity reaches a local maximum at a radius of $\sim$15 kpc, drops
slightly, and then rises to the inclination-corrected value of
304\kms, remaining relatively flat and symmetric out to a radius of
$\sim$50 kpc.  The rotation velocity is in close agreement with that
implied by the inclination-corrected 50\% velocity width (310
\kms). 

\subsection{Infrared observations and photometry} 

The 4-m Anglo-Australian Telescope (AAT) was used to obtain $K_{\rm
s}$- and $H$-band images of \gal\ on 2003 April 9, with exposure times
of 660~s and 420~s, respectively, and $1^{\prime\prime}\!\!.8 -
2^{\prime\prime}\!\!.0$ seeing. The final images were made by random
dithering of 11 $K_{\rm s}$-band images and 7 $H$-band images in a
$\pm$30\arcsec\ box. The positions were calibrated using the {\sc
koords} tool in Karma (Gooch 1996), with a DENIS (Epchtein 1997) image
of the same field serving as the reference. The DENIS positions are
accurate to better than 1\arcsec.

To obtain accurate photometry of the three galaxies in the crowded
fields, we first measured and subtracted all of the stars in the
field using the IRAF\footnote{IRAF is distributed by the National
Optical Astronomy Observatories, which are operated by the Association
of Universities for Research in Astronomy, Inc., under cooperative
agreement with the National Science Foundation.} package {\sc killall}
(see Woudt et al. 2005 for a detailed description). The resulting
star-subtracted images are shown in Fig.~8 and Fig.~9. We then used
the IRAF task {\sc ellipse} to derive the integrated magnitudes for
all three galaxies; they are shown in Fig. 10 and listed in Table 4
together with data from 2MASS and DENIS.  We calibrated the AAT images
by comparing the AAT point sources, measured as described above, with
sources from the 2MASS Point Source Catalog\footnote{This publication
makes use of data products from the Two Micron All Sky Survey, which
is a joint project of the University of Massachusetts and the Infrared
Processing and Analysis Center/California Institute of Technology,
funded by the National Aeronautics and Space Administration and the
National Science Foundation.}.

\begin{figure*}
\begin{tabular}{c}
   \mbox{\psfig{file=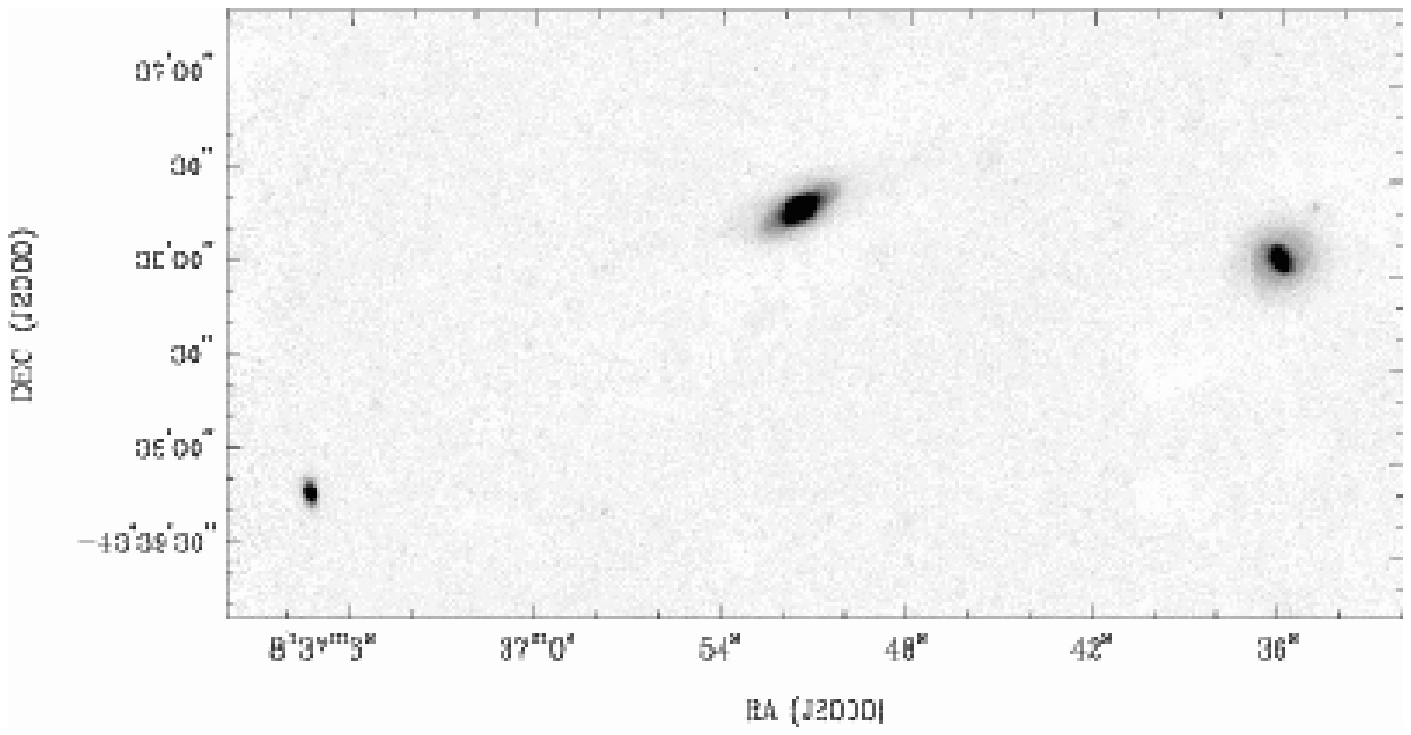,width=15.2cm,angle=0}} \\
   \mbox{\psfig{file=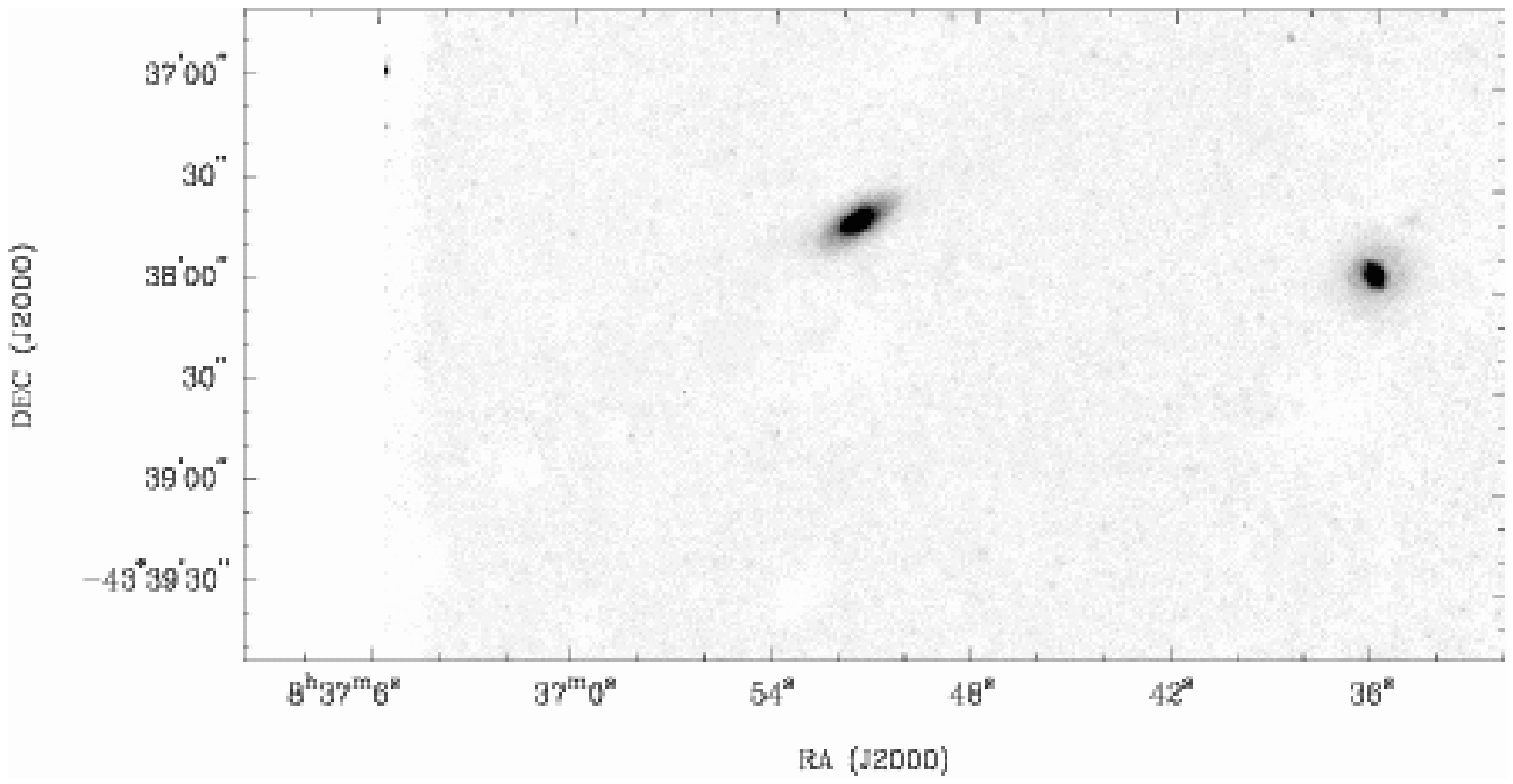,width=15.2cm,angle=0}} \\
\end{tabular}
\caption{Star-subtracted AAT $K_{\rm s}$-band image (top) and $H$-band image
    (bottom) of \gal\ and neighbouring galaxies.}
\end{figure*}

\begin{figure*}
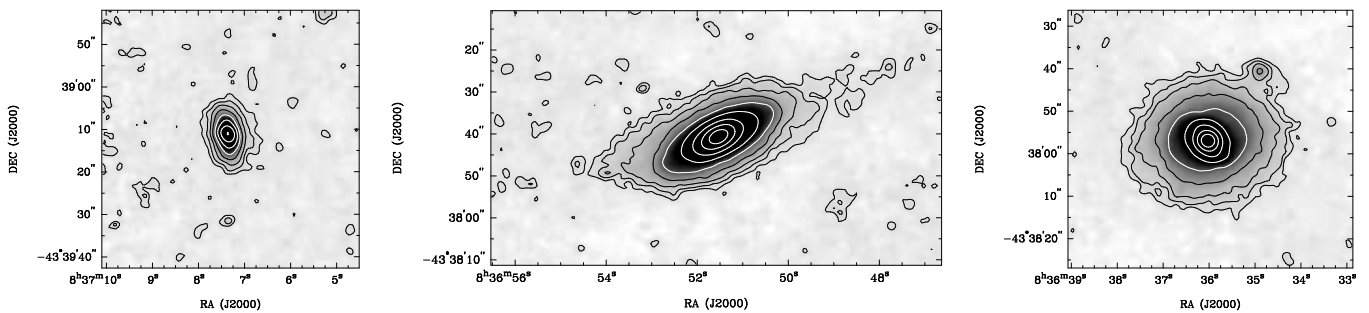

\begin{tabular}{ccc}
   \mbox{\psfig{file=fig9a.eps,height=4cm,angle=-90}} &
   \mbox{\psfig{file=fig9b.eps,height=4cm,angle=-90}} &
   \mbox{\psfig{file=fig9c.eps,height=4cm,angle=-90}} \\
\end{tabular}
\caption{Star-subtracted AAT $K_{\rm s}$-band images of \GGG\ (left), 
\gal\ (middle) and \GG\ (right). The contour levels are 30, 80, 160,
 320, 640, 1280, 2560, 5120, and 10240 units. The images were
 convolved with a 1\farcs5 Gaussian.}
\end{figure*}

\begin{figure*}
   \mbox{\psfig{file=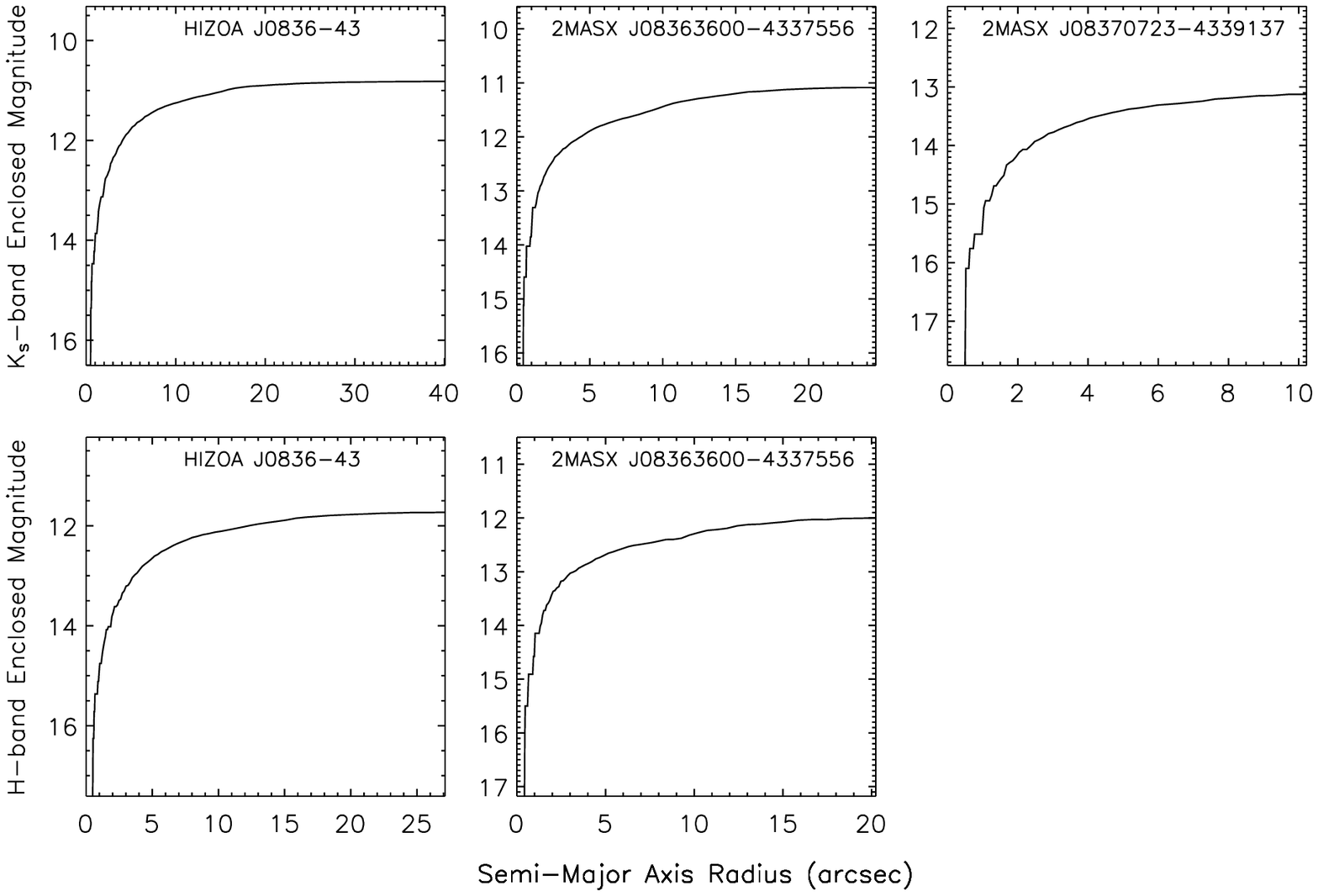,width=16cm}}
\caption{Enclosed AAT $K_{\rm s}$-band (top) and $H$-band (bottom) elliptical 
   magnitudes as a function of semi-major axis radius (see also Table~4). 
   Magnitudes have not been corrected for extinction.}
\end{figure*}

The radial surface brightness profiles are shown in Fig.~11. As in
Beijersbergen et al. (1999) and Galaz et al. (2002), we simultaneously
fit the surface brightness data by two exponential profiles, taken to
be the bulge and disk components of the galaxy:
\begin{equation}
\mu(r) = \mu_{\circ} + 1.086 \times {r \over h}~~{\rm mag~arcsec^{-2}}
\end{equation} 
The surface brightness profile parameters are given in Table~3. We
used a geometric (multiplicative) step when measuring the surface
brightness profile because such a step does not oversample the outer
regions, where little flux is detected. Changing to a linear step
gives more weight to the disk component and tends to brighten the disk
central surface brightness by an average of 0.7 mag and decrease the
disk scale height by $\sim$1\arcsec; the change in the bulge
parameters is less significant. A linear step also changes the central
extinction-corrected ($H$-$K_{\rm s}$) colours of the bulge and the
disk, from 0.41 and 0.50, to 0.44 and 0.34, respectively. When
corrected for the inclination of the galaxy assuming no optical depth,
the face-on central surface brightness of the disk and bulge
components, $\mu_{\rm fo} = \mu_{\rm i} - 2.5$ log(cos$i$), becomes
fainter by 1 mag. This correction is almost exactly cancelled by the
$K$-band extinction correction.

\begin{figure*}
   \mbox{\psfig{file=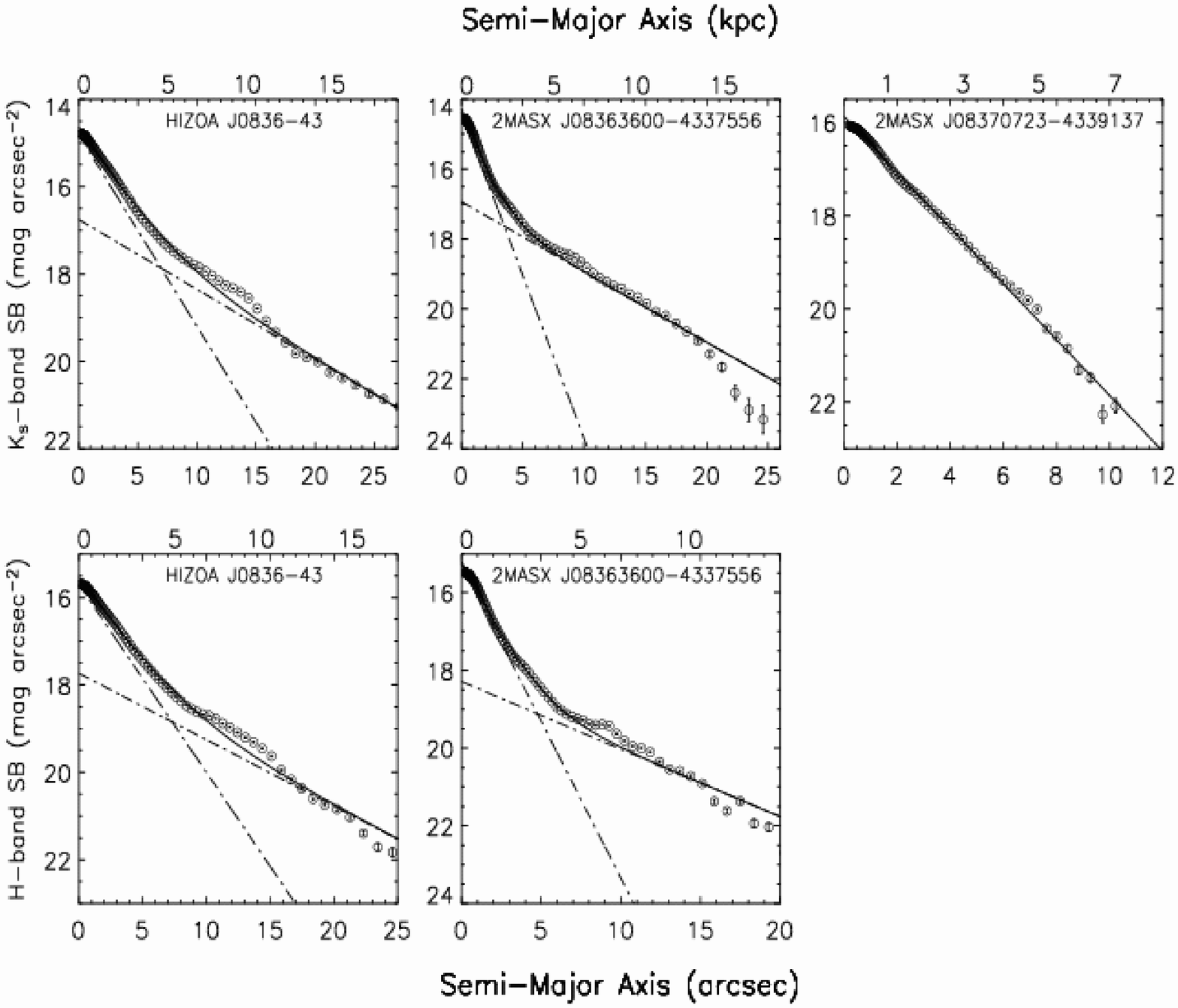,width=14.5cm}}
\caption{Observed AAT $K_{\rm s}$- and $H$-band surface brightness (SB) 
   profiles. The profiles of \gal\ and 2MASX~J08363600--4337556 were 
   simultaneously fit by two exponential profiles, given by the dot-dashed 
   lines. The solid line shows the sum of the two fits. The profile of the 
   third galaxy, 2MASX~J08370723--4339137, was fit by a single exponential 
   profile.}
\end{figure*}

The bulge of \gal\ has a central observed surface brightness of
$\mu_{\circ}$ = 14.81 (15.70) mag\,arcsec$^{-2}$ in the $K_{\rm s}$-
($H$~-) band images and a scale length of $h \sim 2.5\arcsec$ or 1.7
kpc. The disk of \gal\ has a central observed surface brightness of
$\mu_{\circ}$ = 16.75 (17.74) mag\,arcsec$^{-2}$ and a scale length of
$h \sim 7\arcsec$ or 4.7 kpc.  The bulge-to-disk ratio, $B/D$,
calculated from the surface brightness profile fits, is 0.80 (0.83) in
the $K_{\rm s}$- ($H$~-) band.

The integrated infrared magnitudes are given in Table~4, together with
the infrared properties measured in 2MASS, and DENIS.  The formal
statistical error on all AAT magnitudes listed in Table~4 is
$\sim$0.01 mag; this error does not include potential systematic
errors introduced during image processing.  The inclination of \gal,
measured from the surface brightness fits to the star-subtracted
images, $i \sim 66$\degr, was calculated assuming an intermediate
value of $p=c/a=0.15$ (see Haynes \& Giovanelli 1984).

\section{The Galaxy HIZOA J0836--43} 

The overall properties of the galaxy \gal, derived from our \HI\ and
infrared data, are summarised in Table~2. Below, we discuss the \HI\
and dynamical mass of \gal, along with its star formation rate,
morphology, NIR colours, location on the Tully-Fisher relation, and
environment.

\subsection{The \HI\ and dynamical mass} 

The \HI\ flux density of \gal\ as obtained from the Parkes HIZOA data is 
\FHI\ = $13.2 \pm 1.3$ Jy\kms, similar to the value of $14.5 \pm 0.7$ Jy\kms\ 
measured with the ATCA in 750D array (see Table~1), suggesting that
little (if any) \HI\ emission was resolved out by the interferometer
in this configuration. We adopt the ATCA value to estimate an \HI\
mass of \MHI\ = 7.5 ($\pm$0.3) $ \times 10^{10}$~\msun. Assuming the
other two galaxies in the field (see Figs.~3 and 8) lie at the same
distance as \gal, we estimate an upper limit to their individual \HI\
masses of $\sim5 \times 10^9$~\msun.

The \HI\ diameter of \gal\ is at least 3\arcmin, corresponding to 120
kpc. Our analysis of the \HI\ radial density profile indicates a
face-on radius of 66 kpc, at a surface density of
1~\msun\,pc$^{-2}$. Using the \HI\ radius and the
inclination-corrected rotational velocity of $v^i_{\rm rot}$ =
304\kms, we derive a total enclosed dynamical mass of \Mdyn\ = $1.4
\times 10^{12}$~\msun. The ratio of \HI\ mass to dynamical mass is
0.053.

\subsection{Star formation rate} 
 
The 20-cm radio continuum data reveal a slightly extended source
coincident with \gal\ (see Fig.~3). The source has a total flux
density of $\sim$24 mJy, a deconvolved size of $11\arcsec \times
7\arcsec$, and a major axis position angle of 120\degr. The centre
position of the radio continuum source is $\alpha,\delta$(J2000) =
08:36:51.54, --34:37:41.5 ($\pm$0.1\arcsec). Using the relation
calibrated by Bell (2003), we calculate a star formation rate (SFR) of
35~\msun\,yr$^{-1}$. Assuming that $M_{\rm gas} = 1.4 \times$\MHI, the
gas consumption time, $M_{\rm gas}$/SFR, for \gal\ is 3 Gyr and is
therefore quite typical for spiral galaxies (Kennicutt et al. 1994).

The far-infrared (FIR) 60 \micron\ flux density of \gal\ from the High
Resolution Infrared Astronomical Satellite (IRAS) Atlas (Cao et
al. 1997) confirms that star formation, and not AGN activity, is
responsible for the radio continuum emission.  Although \gal\ is
clearly detected at 60 \micron\ (and marginally detected at 100
\micron), source confusion prevents measurement of an accurate flux.
We instead place conservative upper and lower limits on the 60
\micron\ flux density by summing the galaxy emission inside an
elliptical aperture and subtracting both the low and high extremes of
the local sky; we measure a 60 \micron\ flux density of $\sim
1.5-4.8$~Jy.

From the FIR 60 \micron\ flux density limits, the radio/infrared
correlation for star-forming galaxies and radio-quiet AGN (Yun, Reddy,
and Condon 2001) predicts radio luminosities that are a factor of 0.5
-- 1.6 times that observed. The observed radio luminosity therefore
exceeds the expected luminosity by at most a factor of 2, falling far
below the factor of 5 required for 'radio-excess' AGN (Yun et
al. 2001, Drake et al. 2003), and confirming that the radio continuum
emission is dominated by star formation.

While there is no sign of any radio continuum emission at the position
of the galaxy 2MASX~J08363600--4337556, west of \gal, the smaller
galaxy to the south-east of \gal, 2MASX J08370723--4339137, is
detected with a 20-cm flux density of $\sim$1 mJy (see bottom panel of
Fig 3). The rms in the continuum image is $\sim$0.1 mJy~beam$^{-1}$.

\subsection{Morphology and colours} 

The morphology of \gal\ remains a puzzle. The \HI\ observations show a
rapidly rotating, very gas-rich galaxy with --- as seen in the
previous section --- a high star formation rate, indicative of an Sb
or Sc galaxy (Roberts \& Haynes 1994, Kennicutt 1992).  In addition,
the \MHI/\Mdyn\ ratio of \gal\ appears typical for Sb--Sc galaxies,
but would be within the observed range of values for any galaxy type
(Roberts \& Haynes 1994). Visual examination of the AAT $K_{\rm
s}$-band image of \gal\ (Fig.~3), however, and the analysis of its
surface brightness profile (Fig.~11 and Table~3), reveal a distinct
bright bulge ($\sim$2 kpc) and an extended smooth disk ($\sim$5 kpc)
that can be traced along the major axis out to a radius of $\sim$13.4
kpc (at 20 mag\,arcsec$^{-2}$). Comparison with the NIR Galaxy
Morphology Atlas of Jarrett (2000) leads to the conclusion that \gal,
with its dominant bulge and smooth bulge-to-disk transition, is more
likely to be a bright S0 to Sa galaxy than an Sb or Sc.  NIR
extinction from the Galactic Plane, however, is likely to obscure the
disk, causing \gal\ to appear anomalously early in type. Using the NIR
positions of \gal\ and its companions, we identify highly obscured
optical counterparts in the DSS2 $R$-band images, but the image
resolution is insufficient to provide any details about their
morphologies.

The extinction-corrected NIR colours of \gal, presented in Table 5,
are consistent with expectations for normal bright galaxies ($J-K_{\rm
s}$ = 1\fm0, $H-K_{\rm s}$ = 0\fm27, and $J-H$ = 0\fm73), but provide
little morphological information as only late-type spirals (Sd -- Sm)
deviate from the overall mean (Huchra et al. 2000; Jarrett et
al. 2003). The AAT $(H-K_{\rm s})^o_{5\arcsec}$ values of $0\fm39$ and
$0\fm34$ for \gal\ and its larger companion, \GG, lie on the high side
of the range for normal 2MASS galaxies; the same colours determined
from the 2MASS magnitudes lie on the low side ($0\fm23$ and
$0\fm15$). We use the 5\arcsec\ colours instead of the total colours
because the former are less likely to be affected by extinction and
star-subtraction. The $(J-K_{\rm s})$ colour of the small galaxy,
\GGG, lies well above the mean, which might be indicative of a late
type spiral. In addition, comparison of DENIS central and mean
$(I-K_{\rm s})$ colours suggests that the total stellar populations of
\gal\ and \GG\ are bluer than that of the bulges that dominate the
light within $7\arcsec$.

As shown by Jarrett et al. (2003), NIR effective surface brightnesses
brighten with decreasing morphological type, $T$.  After extinction
correction, the mean effective surface brightnesses listed in Table 2
suggest that \gal\ is of early type, in the range E (elliptical) to
Sa, although the H and K$_{\rm s}$ surface brightnesses lie within the
rms uncertainties for types as high as Sb. The face-on companion \GG\
also seems of early type, in the range E to Sa, and the third small
companion \GGG\ seems to be a spiral between Sab and Sc, in agreement
with its S-shaped structure and the indication of star formation (see
the bottom panel of Fig.~3).

As a final test of morphology, we compare the central surface
brightness in the $J$- and $K_{\rm s}$-band with the total $K_{\rm s}$
magnitude. According to Jarrett (2000), SA0/S0 galaxies follow a
linear relation of the form $J_{5"} = 0.8 K_{\rm s} +8.8$, and $K_{5"}
= 0.8 K_{\rm s} + 7.8$. Spiral galaxies of later type ($T$) follow the
same slope but with an increasing offset towards fainter magnitudes
with increasing $T$, while ellipticals generally lie below this
regression. The $J_{5"}$ and the $K_{5"}$ relations again suggests
that the massive spiral galaxy \gal\ is an S0 (lenticular) or Sa
galaxy. The larger companion, \GG, also seems of early type. For the
small spiral-like galaxy, \GGG, the test however, also indicates an
early type which it clearly is not. This may be due to the fact that
this galaxy is very small and hardly resolved; the above relation is
valid only for well resolved galaxies.

The NIR morphology, colours, and magnitudes therefore indicate that
\gal\ may be an S0 or Sa galaxy, although we caution that extinction
is likely to obscure the disk, causing the bulge to dominate the
photometry and therefore bias our results towards early morphological
types.  Given its enormous \HI\ content, it may at first seem
contradictory that \gal\ could be a lenticular galaxy.  However, some
lenticulars are known to contain a large amount of neutral gas.  Such
galaxies include low surface brightness lenticulars, of which the
edge-on S0 galaxy NGC~5084 (Gottesman \& Hawarden 1986) is a typical
example (see also O'Neil et al. 2004), as well as high surface
brightness lenticulars (Eder et al. 1991). For many S0 galaxies --- in
particular polar ring galaxies --- the gas seems to be due to recent
accretion events.  Other \HI-rich lenticulars (e.g., NGC~5084),
however, have \HI\ disks that are aligned with their optical
counterparts as well as similar gas and stellar kinematics, suggesting
that the \HI\ is intrinsic in origin.

\subsection{NIR Tully-Fisher relation} 

In order to test whether \gal\ lies on the Tully-Fisher relation, we
compare our data to the NIR ($R, I, H, K_{\rm s}$) Tully--Fisher
relations as calibrated by Macri (2001). These relations, calibrated
for both total and isophotal magnitudes (20~mag/arcsec$^2$), exhibit
slopes with small dispersions ($0\fm19 - 0\fm22$).

We list in Table 6 the absolute magnitudes estimated from Macri's
Tully-Fisher relations (using the inclination-corrected 20\% velocity
width, $\log w_{20}^i = 2.82$) and the extinction-corrected, absolute
magnitudes obtained from AAT, 2MASS and DENIS photometry\footnote{The
internal extinction was corrected according to Masters et al. (2003)
using $A_\lambda = c_\lambda \log (a / b)$, with $c_{\lambda} = 1.16,
0.39, 0.26$ for $I, J, K_{\rm s}$, respectively, where $(a / b)$ is
the major-to-minor axis ratio.}. The offsets between the predicted and
measured magnitudes, as well as the slopes and intercepts of the
regressions $M = a + b (\log w_{20}^i - 2.5)$, are also listed. For
all bands, \gal\ lies on the NIR TF relation, never deviating by more
than $1\sigma$, suggestive of a normal galaxy at the extreme high-mass
end of the NIR TF relation.

\subsection{The environment} 

Does \gal\ form part of a cluster or a group that allowed it to grow
to its current dimensions through the accretion of or merging with
satellite galaxies, or does it reside in a low-density environment
where it remained undisturbed by galaxy interactions?  Unfortunately,
the study of the environment of \gal\ is made difficult by its
location behind the plane of the Milky Way and behind the Vela
SNR. With optical extinctions of $\sim$10 magnitudes, the detection of
even the intrinsically most luminous galaxies at the distance of \gal\
is nearly impossible. Although NIR surveys are less affected by
extinction than optical surveys, they are hindered by
confusion. Galaxy recognition fails completely in regions where the
density of stars with $K \le 14\fm0$ surpasses $\log N = 4.00
(\Box)^\circ$, and \gal\ (with $\log N = 3.96$) lies extremely close
to this so-called ``NIR Zone of Avoidance''(see Kraan-Korteweg 2005).

Nevertheless, we attempt to explore the environment of \gal\ by
studying all extended sources in the 2MASX catalogue that lie within
(1) a square degree around \gal\ and (2) within a circle of radius
$1\fdg2$, the equivalent of an Abell radius at the distance of \gal.
With the exception of two small pockets of extreme dust, the
extinction $A_K$ in this area ranges from $0\fm6$ to
$1\fm4$. Subtracting the latter from the nominal 2MASS completeness
limit of $K_s \le 13\fm5$ (Jarrett 2000), we expect our extended
source search to be complete to $K_S^\circ \la 12\fm1$ in
extinction-corrected magnitude, ignoring confusion. To exclude
Galactic sources from this sample, we visually inspect all 2MASX
objects brighter than $K_S^\circ \le 12\fm0$.

In the central square degree, there are 14 2MASX extended sources with
$K_S^\circ \le 12\fm0$. Of these, one was rejected, leaving 10 certain
galaxies and 3 possible galaxies. The galaxy density of the studied
region is therefore at least a factor of 5 higher than the mean
density of 2MASS galaxies with $K_S \le 12\fm0$ and $|b| \ge
25^\circ$, 1.8 gal/$(\Box)^\circ$ (Jarrett 2004; Jarrett 2005, priv.
comm.), although the redshift of the overdensity remains
undetermined. Within the Abell radius, 29 extended sources have
$K_S^\circ \le 12\fm0$; of these, six were rejected, leaving 18
certain galaxies plus 5 possible galaxies. The resulting galaxy
density in the Abell radius therefore also exceeds the mean, by a
minimum factor of 2--3.

To investigate the effect of confusion on the overdensity, we repeated
the study for completeness limits of $K_S^\circ \le 11\fm5$ and
$11\fm0$. The mean 2MASX densities (away from the Galactic Plane) are
0.85 and 0.43 gal/$(\Box)^\circ$, respectively. Although the
statistics are poorer, we find galaxy densities in the square degree
around \gal\ of $7 - 10$ and $5 - 8$ gal/$(\Box)^\circ$, respectively,
indicating even higher overdensities than for the $K_S^\circ \le
12\fm0$ completeness limit. The overdensity within the Abell radius
remains approximately the same.  It therefore appears that \gal\ lies
in an overdense region --- at least projected on the sky --- with the
galaxy density increasing in the central square degree and becoming
more prominent for the brightest objects.

Although we do not have redshifts for the individual 2MASS galaxies,
we estimate the redshift of the deduced galaxy overdensity, and its
relation to the redshift of \gal, $z = 0.036$, by considering the
large-scale structure of 2MASS galaxies in several photometric
redshift intervals, as determined by Jarrett (2004).  Prominent in the
slice $0.04 < z < 0.05$, but also visible in the $0.03 < 0.04$ slice,
is a filamentary supercluster-like feature that crosses the Galactic
Plane at the position of \gal\ and that points towards the Shapley
concentration. This feature could explain the projected overdensity of
galaxies around \gal.  In addition, a large fraction of the 2MASS
galaxies seem to be of similar size and appearance as \gal\ (taking
absorption effects into account), and are therefore likely to lie at
similar redshifts.

Finally, we checked for \HI-rich neighbours to \gal\ by observing a
$2\degr \times 2\degr$ area around it with the Parkes multibeam system
in 2004 November. The integration time was $4 \times$ longer than for
the deep Parkes \HI\ ZOA survey, and $20 \times$ longer than for
HIPASS.  The central velocity was set to 10\,000\kms; the 64 MHz
bandwidth give velocity coverage from 3300\kms\ to 16\,700\kms. A
careful inspection of the deep \HI\ data cube did not reveal any
additional galaxies.  Given an rms of $\ga$3 mJy, the 5$\sigma$
\HI-mass limit for a detection at $D$ = 148 Mpc is $\ga1.5
\times 10^{10}$\msun\ (assuming a top-hat with a velocity width of
200\kms). Areas with strong continuum emission are adversely affected
by a baseline ripple which increases the noise by a factor 2--3.

\section{Comparison with other \HI-massive galaxies} 

The four most \HI-massive galaxies known are \gal, UGC\,4288 (O'Neil
et al. 2004), UGC\,1752 (Matthews et al. 2001), and, of course,
Malin\,1 (Bothun et al. 1987). The latter three are low-surface
brightness (LSB) galaxies at luminosity distances of 433, 249 and 351
Mpc, respectively. Their \HI\ properties are summarised in Table~7.

The galaxy UGC\,4288 was classified as Sdm by Schombert \& Bothun
(1988) and as S0 by O'Neil et al. (2004). Optical images show a bright
bulge and faint spiral structure, suggesting that UGC\,4288 is likely
to be of late type. The galaxy UGC\,1752, which was classified as SAcd
by de Vaucouleurs et al. (1991), looks remarkably similar to
UGC\,4288, suggesting the latter is also of late type. Malin\,1 has a
prominent bulge, with the properties of a bright elliptical galaxy,
and a low-surface brightness disk that is barely detectable. The
galaxy \gal\ is highly obscured in the optical and its morphological
type is unknown.  While our analysis of its NIR colours slightly
favour an early-type morphology, extinction is likely to obscure the
disk, causing the measurements to be dominated by the prominent bulge.
The surface brightness analysis gives $B/D \sim 0.8$, a value slightly
higher than that seen in LSBs ($B/D = 0.56 \pm 0.07$; Sprayberry et
al. 1995), but roughly consistent with that of spiral galaxies ($B/D =
0.61 \pm 0.20$; Kent 1985).  The extinction- and inclination-corrected
central $K_{\rm s}$-band surface brightness of the disk
($\mu_{\circ}$=16.9~mag~arcsec$^{-2}$) falls well short of the typical
LSB cut, 18.0~mag~arcsec$^{-2}$, indicating that unlike the other
\HI-massive galaxies listed above, \gal\ is not an LSB galaxy.


The \HI\ diameters of all four galaxies are huge: Malin\,1
($\sim$2\farcm5 or 220 kpc), UGC\,1752 ($\sim$2\arcmin\ or 130 kpc),
and \gal\ ($\sim$3\arcmin\ or 120 kpc). No \HI\ diameter has been
measured for UGC\,4288 (yet) but its optical extent of $\sim$1\arcmin\
(or 100 kpc) is likely to be a lower limit. All four galaxies also
have high rotational velocities; while their inclination angles are
rather uncertain (Malin\,1: $i \approx 45$\degr, UGC\,1752: $i \approx
0$\degr; UGC\,4288 $i \approx 40$\degr; \gal: $i \approx 66$\degr), we
deduce rotational velocities of $\ga$200--300\kms\ from the velocity
width of their \HI\ profiles (see Table~7). We note that UGC\,4288,
UGC\,1752, and Malin\,1 appear to lie in very low density
environments. Their lack of tidal interactions may have played a role
in preserving such large \HI\ masses and diameters.

It therefore appears that the four \HI-massive galaxies discussed
above are similar in their extreme \HI\ masses, large disks, and high
rotational velocities.  It is possible, however, that \gal\ differs
from the others in its high SFR, high surface brightness disk,
potentially overdense environment, and morphology.  A more detailed
comparison of these galaxies will follow in Koribalski et al. (in
preparation).

\section{Conclusions} 

HIZOA J0836--43 is a remarkable galaxy. Its enormous \HI\ mass (\MHI\
= 7.5 ($\pm$0.3) $\times 10^{10}$~\msun), large \HI\ velocity width
($w_{20} = 610\pm~9$\kms), and paucity of optical information due to
its location in the Zone of Avoidance raises numerous questions about
its environment, morphology and overall kinematics. ATCA \HI\
observations reveal a very extended, highly-inclined \HI\ disk (see
Fig.~2) and a surprisingly regular velocity field. We measure an
\HI\ diameter of at least 3\arcmin\ which corresponds to a physical
size of $\sim$120 kpc. For comparison, the \HI\ diameter of Malin\,1,
the largest disk galaxy known so far, is $\sim$220 kpc (Pickering et
al. 1997, Matthews et al. 2001).

The \HI\ distribution of \gal\ is slightly asymmetric (with more \HI\
gas on the eastern side of the galaxy) and reveals a central
depression. It also shows a gradually changing major axis position
angle which decreases from $PA \sim 290\degr$ in the inner region to
nearly 270\degr\ in the outermost region. This change, which is most
pronounced on the eastern side of the galaxy, is indicative of a
gentle warp.

The \HI\ velocities observed in \gal\ range from $\sim 10400$ to
11000\kms, indicating an inclination-corrected rotation velocity of
304\kms\ and consequently, a large dynamical mass and dark matter
content. \gal\ falls on the NIR Tully-Fisher relationship, indicating
that it is a normal rotating galaxy. We note that the equal velocity
contours are nearly straight and do not resemble a typical spider
diagram. A detailed analysis of the galaxy kinematics is limited by
the relatively low angular resolution.

NIR data from the AAT indicate that \gal\ has a both a prominent bulge
and an extended disk, and NIR colours point to an early type (S0-Sa)
morphology, although colours from AAT, 2MASS, and DENIS may be
dominated by the bulge component due to obscuration of the disk.  The
high star formation rate measured from the radio continuum and IRAS
data, $SFR \sim 35$~\msun~yr$^{-1}$, suggests a later morphological
type of Sb or Sc. A study of the local environment of \gal, using
sources from the 2MASX extended source catalogue, indicate that \gal\
may lie in an overdense region, possibly part of a supercluster-like
filament.

There are currently very few galaxies known with \HI\ masses of
$\sim10^{11}$\msun; these are \gal, UGC\,1752, Malin\,1, and UGC\,4288
(in order of increasing distance; see Table~7). Accurate \HI\ mass
estimates can be difficult for such distant galaxies because of flux
calibration uncertainties and low signal to noise. In addition,
single-dish \HI\ spectra are often fitted with multiple-order
baselines which result in large flux uncertainties due to the very
wide \HI\ velocity profiles of these supermassive galaxies. We have
obtained deep \HI\ spectra of the LSB galaxies Malin\,1 and UGC\,4288
with the Parkes 64-m telescope and will report the results in
Koribalski et al. (in preparation) together with a comprehensive
comparison of the most \HI-massive galaxies known from the
literature. Constraining the number of massive galaxies in the
Universe is necessary if we are to better understand the high-mass end
of the galaxy mass and luminosity functions, and studying the
properties of such galaxies provides a test of current galaxy
formation scenarios in their most extreme limits.

\section*{Acknowledgments}
We are grateful to Patrick Woudt for his help with the
star-subtraction in the AAT $K$- and $H$-band images. We thank Stacy
Mader for obtaining the sensitive Parkes \HI\ data of the $2\degr
\times 2\degr$ area around \gal, and the referee, Greg Bothun, for
helpful comments.  This research has made use of the NASA/IPAC
Infrared Science Archive (2MASS) and the NASA/IPAC Extragalactic
Database (NED), which are operated by the Jet Propulsion Laboratory,
California Institute of Technology, under contract with the National
Aeronautics and Space Administration. RCKK thanks UCT, the SA National
Research Foundation and CONACyT (research grant 40094F) for their
support, and the Australian Telescope National Facility (ATNF, CSIRO)
for their hospitality during her sabbatical.  JLD performed most of
this research while a Fulbright scholar at the ATNF.

\clearpage

\begin{table*} 
\caption{Measured \HI\ properties of the galaxy \gal}
\begin{tabular}{lcccccccc} 
\hline
Observation  &rms  &\HI\ flux density&Heliocentric  &\multicolumn{2}{c}{Velocity width} & Reference \\
             &     &  \FHI\          &velocity, $cz$& $w_{50}$ & $w_{20}$ &     \\
             &[mJy~beam$^{-1}$] &  [Jy\kms]       & [\kms]       & [\kms]   &  [\kms]  &      \\
\hline
Parkes ZOA survey   &6     & $13.2\pm1.3$  & $10706\pm~7$  & $550\pm14$  & $607\pm21$   & here \\
ATCA                &1.2   & $14.5\pm0.7$  & $10689\pm~3$  & $566\pm~6$  & $610\pm~9$   & here \\
HIPASS              &13    & $15.6\pm2.9$  & $10710\pm16$  & $575\pm32$  & $654\pm48$   & HICAT (Meyer et al. 2004) \\ 
\hline
\end{tabular}
\flushleft
{\bf Note:} The uncertainties were calculated following the formulas 
           given in Koribalski et al. (2004) and references therein.
\end{table*}

\begin{table*} 
\caption{A summary of the properties of the galaxy HIZOA J0836--43}
\begin{tabular}{lcc} 
\hline
Property & Value &  Notes\\  
\hline
centre position, $\alpha,\delta$(J2000)                       & $08^{\rm h} 36^{\rm m} 51.54^{\rm s}$, --43\degr 37\arcmin 41\farcs5 & (a)\\
heliocentric velocity, \vhel                                  & $10689 \pm 3$\kms          & (b)\\
CMB velocity, \vCMB                                           & $10928 \pm 3$\kms          & $\cdots$ \\
luminosity distance                                           & 148 Mpc                    & (c)\\
\HI\ flux density, \FHI                                       & $14.5 \pm 0.7$ Jy\kms      & (b)\\
\HI\ mass, \MHI\                                              & $7.5 \times 10^{10}$ \msun & (b)\\
position angle, $PA$                                          & $285\degr \pm 15\degr$     & (d)\\
inclination angle, $i$                                        & $66\degr \pm 3\degr$       & (d)\\
intrinsic rotation velocity, $v^i_{\rm rot}$                  & 304\kms                    &$\cdots$ \\
\HI\ radius (at 1 \msun\,pc$^{-2}$)                           & 66 kpc                     & (e)\\
total dynamical mass, \Mdyn                                   & $1.4 \times 10^{12}$ \msun & $\cdots$\\
absolute magnitude, $M_{K_{\rm s}}^c$                         & --25.03 mag                & (d,f,g)\\
infrared colour, $(H-K_{\rm s})$                               & 0.43                       & (d,f,g)\\
\HI\ mass to light ratio, \MHI\ / $L_{K_{\rm s}}$             & 0.35 \msun/\lsun           & (f,g)\\
\HI\ mass to total mass ratio, \MHI\ / \Mdyn                  & 0.05                       & $\cdots$ \\
infrared radius, ($\mu_{K_{\rm s}}^c$ = 20 mag arcsec$^{-2}$) & 13.4 kpc                   & (d,f)\\
bulge scale length                                            & 1.7 kpc                    & (d)\\
disk scale length                                             & 4.7 kpc                    & (d)\\
star formation rate (SFR)                                     & 35 \msun\,yr$^{-1}$        & (a)\\
\hline
\end{tabular}

\flushleft
{\bf Notes:}
(a) from the ATCA 20-cm radio continuum data;
(b) from the ATCA 750D array \HI\ line data;
(c) assuming $H_0$ = 75\kms\,Mpc$^{-1}$ (see \S~3.1.1);
(d) from the AAT $H$- and $K_{\rm s}$-band data;
(e) when viewed face-on;
(f) corrected for Galactic extinction (Schlegel et al. 1998);
(g) the absolute magnitude of the Sun in $K$-band is $M_{K,\odot} = 3.31$ mag
    (Colina \& Bohlin 1997).
\end{table*}

\begin{table*} 
\caption{Observed Surface Brightness Profile Parameters}
\begin{tabular}{llccccc} 
\hline
                         &                & 
                         & \multicolumn{2}{c}{Bulge Parameters}
                         & \multicolumn{2}{c}{Disk Parameters}  \\
Source                   & Band           & 
$\mu_{\circ}$            & $\mu_{\circ}$  & h 
	                 & $\mu_{\circ}$  & h \\
                         &                & 
[mag\,arcsec$^{-2}$]     & [mag\,arcsec$^{-2}$] & [$^{\prime\prime}$] 
		         & [mag\,arcsec$^{-2}$] & [$^{\prime\prime}$] \\
\hline
HIZOA J0836--43          & $H$         & 15.55 & 15.70 & 2.53  & 17.74 & 7.15 \\
                         & $K_{\rm s}$ & 14.65 & 14.81 & 2.48  & 16.75 & 6.79 \\
2MASX J08363600--4337556 & $H$         & 15.20 & 15.27 & 1.34  & 18.29 & 6.25 \\
                         & $K_{\rm s}$ & 14.27 & 14.37 & 1.16  & 16.93 & 5.40 \\
2MASX J08370723--4339137 & $K_{\rm s}$ & 15.87 & 15.87 & 1.81  & $\cdots$ & $\cdots$ \\
\hline
\end{tabular}
\flushleft 
{\bf Note:} Magnitudes have not been corrected for foreground
extinction, values of which are listed in Table 4.

\end{table*}

\begin{table*} 
\caption{Observed Infrared Parameters}
\begin{tabular}{llclrrcccc}
\hline
Source & 
Band & 
A$_{\lambda}$   & 
Telescope       & 
r$_{\rm{20}}$   & 
r$_{\rm eff}$   & 
m$_{\rm tot}$   & 
m$_{20}$        &
m(r$<$5\arcsec) & 
SB(r$_{\rm eff}$)\\ 
&      
& 
[mag]&           
 &
[$^{\prime\prime}$] &
[$^{\prime\prime}$] &               
&                 
&                    
&
[mag\,arcsec$^{-2}$]\\

\hline
HIZOA J0836--43           & $I$           & 4.40    & DENIS   &$\cdots$ &$\cdots$ & 16.59 $\pm$ 0.10 & $\cdots$          & $\cdots$         & $\cdots$ \\    
                          & $J$           & 2.05    & 2MASS   & 11.2    &  7.7    & 13.16 $\pm$ 0.13 & 13.25 $\pm$ 0.08  & 13.99 $\pm$ 0.05 & 18.75    \\  
                          &               &$\cdots$ & DENIS   &$\cdots$ &$\cdots$ & 13.17 $\pm$ 0.04 & $\cdots$          & $\cdots$         & $\cdots$ \\    
                          & $H$           & 1.31    & AAT     & 16.1    &  6.0    & 11.73 $\pm$ 0.01 & 11.84 $\pm$ 0.01  & 12.39 $\pm$ 0.01 & 17.77    \\  
                          &               &$\cdots$ & 2MASS   & 20.3    & 10.8    & 11.44 $\pm$ 0.06 & 11.81 $\pm$ 0.06  & 12.55 $\pm$ 0.03 & 17.71    \\  
                          &  $K_{\rm s}$  & 0.83    & AAT     & 20.2    &  6.3    & 10.82 $\pm$ 0.01 & 10.90 $\pm$ 0.01  & 11.52 $\pm$ 0.01 & 17.02    \\  
                          &               &$\cdots$ & 2MASS   & 17.8    &  9.6    & 10.84 $\pm$ 0.06 & 11.18 $\pm$ 0.06  & 11.84 $\pm$ 0.02 & 16.86    \\  
                          &               &$\cdots$ & DENIS   &$\cdots$ &$\cdots$ & 11.02 $\pm$ 0.10 & $\cdots$          & $\cdots$         & $\cdots$ \\   
\hline														                         
2MASX J08363600--4337556  & $I$           & 4.27    & DENIS   &$\cdots$ &$\cdots$ & 17.04 $\pm$ 0.08 & $\cdots$          & $\cdots$         & $\cdots$ \\    
                          & $J$           & 1.99    & 2MASS   & 6.7     &  5.1    & 13.37 $\pm$ 0.14 & 13.43 $\pm$ 0.10  & 14.10 $\pm$ 0.05 & 18.81    \\  
                          &               &$\cdots$ & DENIS   &$\cdots$ &$\cdots$ & 13.63 $\pm$ 0.03 & $\cdots$          & $\cdots$         & $\cdots$ \\    
                          & $H$           & 1.27    & AAT     & 11.2    &  4.5    & 12.00 $\pm$ 0.01 & 12.22 $\pm$ 0.01  & 12.62 $\pm$ 0.01 & 18.22    \\  
                          &               &$\cdots$ & 2MASS   & 11.6    &  6.6    & 11.84 $\pm$ 0.08 & 12.08 $\pm$ 0.08  & 12.76 $\pm$ 0.04 & 17.84    \\  
                          &  K$_{\rm s}$  & 0.81    & AAT     & 15.6    &  5.4    & 11.08 $\pm$ 0.01 & 11.18 $\pm$ 0.01  & 11.82 $\pm$ 0.01 & 17.76    \\  
                          &               &$\cdots$ & 2MASS   & 13.9    &  6.0    & 11.31 $\pm$ 0.08 & 11.45 $\pm$ 0.08  & 12.15 $\pm$ 0.03 & 17.06    \\  
                          &               &$\cdots$ & DENIS   &$\cdots$ &$\cdots$ & 11.41 $\pm$ 0.06 & $\cdots$          & $\cdots$         & $\cdots$ \\    
\hline														                         
2MASX J08370723--4339137  & $J$           & 1.95    & 2MASS   &$\cdots$ &  2.5    & 15.92 $\pm$ 0.37 & $\cdots$          & 16.33 $\pm$ 0.40 & 19.91    \\  
                          & $H$           & 1.24    & 2MASS   &$\cdots$ &  2.2    & 14.55 $\pm$ 0.25 & $\cdots$          & 14.62 $\pm$ 0.19 & 18.24    \\  
                          & $K_{\rm s}$   & 0.79    & AAT     & 7.3     &  2.7    & 13.13 $\pm$ 0.01 & 13.24 $\pm$ 0.01  & 13.29 $\pm$ 0.01 & 17.50    \\    
                          &               &$\cdots$ & 2MASS   & 5.4     &  3.0    & 13.51 $\pm$ 0.22 & 13.71 $\pm$ 0.14  & 13.77 $\pm$ 0.14 & 17.89    \\      
\hline
\end{tabular}
\flushleft
{\bf Table columns:} 
  (1) Source name;
  (2) NIR band;
  (3) extinction within the given band;
  (4) telescope; 
  (5) r$_{\rm{20}}$ is the radius at 20 mag \,arcsec$^{-2}$;
  (6) r$_{\rm eff}$ is the integrated half-light radius;
  (7) m$_{\rm tot}$ is the total apparent magnitude;
  (8) m$_{20}$ is the 20 mag \,arcsec$^{-2}$ elliptical isophotal magnitude 
  (9) m(r$<$5\arcsec) is the 5\arcsec\ circular aperture magnitude; and
 (10) SB(r$_{\rm eff}$) is the mean surface brightness at r$_{\rm eff}$.
\end{table*}

\begin{table*} 
\caption{NIR colours corrected for galactic foreground extinction}
\begin{tabular}{lrclrclc}
\hline
      &     &       &            & \multicolumn{3}{c}{2MASX}  & 2MASX \\
Colour & \multicolumn{3}{c}{\gal} 
\     & \multicolumn{3}{c}{J08363600--4337556} & J08370723--4339137 \\
      & AAT & 2MASS & DENIS      & AAT & 2MASS & DENIS        & 2MASS \\
\hline
$(I-J)^o_T$	         &$\cdots$ &$\cdots$ & 1.07      &$\cdots$ &$\cdots$ & 1.13      & $\cdots$ \\
$(I-J)^o_{7\arcsec}$     &$\cdots$ &$\cdots$ & 0.64      &$\cdots$ &$\cdots$ & 0.76      & $\cdots$ \\
                         &$\cdots$ &$\cdots$ &$\cdots$   &$\cdots$ &$\cdots$ &$\cdots$   & $\cdots$ \\   
$(I-K_s)^o_T$	         &$\cdots$ &$\cdots$ & 2.00      &$\cdots$ &$\cdots$ & 2.17      & $\cdots$ \\
$(I-K_s)^o_{7\arcsec}$   &$\cdots$ &$\cdots$ & 1.50      &$\cdots$ &$\cdots$ & 1.49      & $\cdots$ \\
                         &$\cdots$ &$\cdots$ &$\cdots$   &$\cdots$ &$\cdots$ &$\cdots$   & $\cdots$ \\   
$(J-K_s)^o_T$            &$\cdots$ & 1.10    & 0.93      &$\cdots$ & 0.88    & 1.04      & 1.25     \\
$(J-K_s)^o_{5\arcsec}$   &$\cdots$ & 0.93    &0.87$^*$   &$\cdots$ & 0.77    &0.72$^*$   & 1.40     \\
                         &$\cdots$ &$\cdots$ &$\cdots$   &$\cdots$ &$\cdots$ &$\cdots$   & $\cdots$ \\   
$(H-K_s)^o_T$            & 0.43    & 0.12    &$\cdots$   & 0.46    & 0.07    &$\cdots$   & 0.59     \\
$(H-K_s)^o_{5\arcsec}$   & 0.39    & 0.23    &$\cdots$   & 0.34    & 0.15    &$\cdots$   & 0.40     \\
                         &$\cdots$ &$\cdots$ &$\cdots$   &$\cdots$ &$\cdots$ &$\cdots$   & $\cdots$ \\   
$(J-H)^o_T $             &$\cdots$ & 0.98    &$\cdots$   &$\cdots$ & 0.81    &$\cdots$   & 0.66     \\
$(J-H)^o_{5\arcsec}$     &$\cdots$ & 0.70    &$\cdots$   &$\cdots$ & 0.62    &$\cdots$   & 1.00     \\
\hline
\end{tabular}
\flushleft
{\bf Note:} $^*$measured at 7\arcsec.
\end{table*} 

\begin{table*} 
\caption{NIR Tully-Fisher comparisons}
\begin{tabular}{lrrlll} 
\hline
Mag        &Intercept$^{\rm a}$ & Slope$^{\rm a}$  &M$_\lambda$(TF)&M$_\lambda$(phot)
                                                        & $\Delta$(TF-phot) \\
\hline
$I_T$	   & $-$21.10 &  $-$8.7 & $-$23.88 & $-$23.92 (DE)  &   +0.04 (DE)  \\
\\
$H_T$      & $-$22.40 & $-$10.0 & $-$25.60 & $-$25.52 (AAT) & $-$0.08 (AAT) \\
           &          &         &          & $-$25.81 (2M)  &   +0.21 (2M)  \\
\\
$K_T$	   & $-$22.66 &  $-$9.9 & $-$25.82 & $-$26.04 (AAT) &   +0.22 (AAT) \\
           &          &         &          & $-$25.90 (2M)  &   +0.08 (2M)  \\
           &          &	        &	   & $-$25.72 (DE)  & $-$0.10 (DE)  \\
\\
$H_{20}$   & $-$22.01 & $-$10.5 & $-$25.37 & $-$25.41 (AAT) &   +0.04 (AAT) \\
           &          &         &          & $-$25.44 (2M)  &   +0.07 (2M)  \\
\\
$K_{20}$   & $-$22.34 & $-$10.4 & $-$25.67 & $-$25.84 (AAT) &   +0.17 (AAT) \\
           &          &         &          & $-$25.56 (2M)  & $-$0.11 (2M)\\
\hline
\end{tabular}
\flushleft
{\bf Notes:}
(a) Intercept and slope were taken from Macri 2001.
\end{table*} 

\begin{table*} 
\caption{\HI\ measurements and derived properties of some of the most 
         \HI-massive galaxies known.}
\begin{tabular}{lcccrrcll}
\hline
\HI\ source     &  \vhel&  $D$ & \FHI &log \MHI&\wfi&\wtw& Ref., comments \\
                & [\kms]& [Mpc]&Jy\kms&[\msun]&    &    &    \\
\hline
UGC\,4288       & 30223 & 433  & 2.54 & 11.05 & 520& 558& 1, Arecibo \\
UGC\,1752       & 17861 & 249  & 5.13 & 10.87 & 388& 429& 2, Nancay \\
HIZOA J0836--43 & 10689 & 148  &14.5  & 10.87 & 566& 610& 3, ATCA image\\
\hline
Malin\,1        & 24784 & 351  & 1.8  & 10.72 & 293& 341& 2, Nancay \\
   "            & 24755 &  "   & 2.5  & 10.86 &    & 322& 4, VLA image \\
   "            & 24705 &  "   & 2.7  & 10.89 & 295& 355& 5, Arecibo \\
   "            & 24745 &  "   & 4.6  & 11.13 & 315& 340& 6, NRAO 43-m\\
\hline
\end{tabular}
\flushleft
{\bf References:} (1) O'Neil et al. 2004, (2) Matthews et al. 2001, (3) this 
 paper, (4) Pickering et al. 1997, (5) Bothun et al. 1987, and (6) Impey \& 
 Bothun 1989.
{\bf Notes:}
 The heliocentric velocity, \vhel, given in Col.(2), was used to calculate the
 luminosity distance, $D$, given in Col.(3) as described in Section~4.
\end{table*}

\end{document}